\documentclass[usegraphicx,usenatbib]{mn2e}
\usepackage{times}
\usepackage{rotating}
\def\mpc{h^{-1} {\rm{Mpc}}}
\def\kms {\rm{km~s^{-1}}}
\def\apj {ApJ}
\def\apjl {ApJL}
\def\apjs {ApJS}
\def\aj {AJ}
\def\aap {A\&A}
\def\mnras {MNRAS}
\def\ub {^{0.1}u}
\def\gb {^{0.1}g}
\def\rb {^{0.1}r}
\def\ib {^{0.1}i}
\def\zb {^{0.1}z}
\def\sz {\scriptsize}
\begin{document}
\title[SDSS DR7: Luminosity function of galaxies in groups]
{Luminosity function of galaxies in groups in the SDSS DR7: 
the dependence on mass, environment and galaxy type}
\author[Zandivarez \& Mart\'inez]
{Ariel Zandivarez\thanks{arielz@oac.uncor.edu} \& H\'ector J. Mart\'inez\\
Instituto de Astronom\'{\i}a Te\'orica y Experimental,
IATE, CONICET$-$Observatorio Astron\'omico, Universidad Nacional de 
C\'ordoba,\\ Laprida 854, X5000BGR, C\'ordoba, Argentina}
\date{\today}
\pagerange{\pageref{firstpage}--\pageref{lastpage}} 
\maketitle
\label{firstpage}
%%%%%%%%%%%%%%%%%%%%%%%%%%%%%%%%%%%%%%%%%%%%%%%%%%%%%%%%%%%%%%%%%%%%%%%%%%%%%%%
%%%%%%%%%%%%%%%%%%%%%%%%%%%%%%%%%%%%%%%%%%%%%%%%%%%%%%%%%%%%%%%%%%%%%%%%%%%%%%%
\begin{abstract}
We perform an exhaustive analysis of the luminosities of galaxies
in groups identified in the Sloan Digital Sky Survey (SDSS) Data Release 7. 
Our main purpose is to perform a detailed study of the Schechter luminosity
function parameters: the characteristic absolute magnitude and the faint
end slope, as a function of group virial mass in order to quantify 
their dependence on environment. We analyse the trends of the
Schechter parameters as a function of group mass for different photometric bands, galaxy populations, 
galaxy positions within the groups, and the group surrounding large scale density. 
We find that the characteristic magnitude brightens and the faint end slope becomes steeper
as a function of mass in all SDSS photometric bands, in agreement with previous results.
From the analysis of different galaxy populations, we observe that 
different methods to split galaxy populations, based on the concentration 
index or the colour-magnitude diagram, produce quite different behaviours in 
the luminosity trends, mainly for the faint end slope. These discrepancies and
the trends with mass mentioned above are explained when analysing the luminosity function of 
galaxies classified simultaneously according to their concentrations and colours.   
We find that only the red spheroids have a luminosity function that
strongly depends on group mass. Late type galaxies, whether blue or red, have
luminosity functions that do not depend on group mass. The intrinsic change in the
luminosity function of spheroids and the varying number contributions of the different types
explain the overall trend of the LF with group mass. 
On the other hand, dividing the galaxy members in the inner and outer regions
of the groups do not introduce a significant difference in the 
Schechter parameter trends, except for the characteristic absolute magnitude
in the high group virial mass range ($\mathcal{M}>1\times 10^{13} M_{\odot}h^{-1}$)
which is an indication of luminosity segregation in massive groups.
Finally, we also analyse the possible influence of the large scale 
surrounding environment on the luminosity function of group galaxies. We find that 
galaxies inhabiting groups at low density regions experience more pronounced 
variations on the Schechter parameters as a function of groups mass, 
while galaxies in groups at high density regions show an almost constant behaviour.
We discuss the possible implications of our findings in the galaxy evolution scenario.
\end{abstract}
%%%%%%%%%%%%%%%%%%%%%%%%%%%%%%%%%%%%%%%%%%%%%%%%%%%%%%%%%%%%%%%%%%%%%%%%%%%%%%%
%%%%%%%%%%%%%%%%%%%%%%%%%%%%%%%%%%%%%%%%%%%%%%%%%%%%%%%%%%%%%%%%%%%%%%%%%%%%%%%
\section{Introduction} 

Since the pioneer works of \citet{einasto74}, \citet{bo78} and \citet{dressler}, 
stating that early-type systems tend to concentrate in high 
density regions, it has become clear that galaxy properties depend on the local environment.
This dependence on environment must hold important information about the history of galaxy 
formation, so it is important to study the connection between the properties of the galaxies 
and their location in the Universe. In particular, a galaxy property that has been widely used 
in the literature is the luminosity, mainly through the analysis of the galaxy luminosity 
function (LF). This function describes the distribution of luminosities of a given population 
of galaxies and, in most cases, the shape of that distribution can be fully described by a 
function with two parameters \citep{schechter76}: the characteristic absolute magnitude 
$M^{\ast}$ and the faint end slope $\alpha$. Although these statistical measurements themselves 
do not give physical explanations about galaxy formation and evolution, they provide important 
constraints on various physical processes involved in such galaxy life stages (e.g.
\citealt{benson03}).

During the 1990's a lot of efforts have been made in order to compute the LF for galaxies in 
different environments such as the field, groups and clusters of galaxies (see for instance 
\citealt{mhg94,lin96,zucca97,lopcruz97,valotto97,ratclif98,mvl98,rauzy98,tren98}). 
However, the advent of large surveys of galaxies, such as the Sloan Digital 
Sky Survey \citep{sdss} and the Two degree Field Galaxy Redshift Survey (2dFGRS, 
\citealt{2df}), allowed for much better determinations of the LF 
\citep{blanton01,blantonlf,blanton05,norb02,madgwick,trentu02,m02,cz03,eke04}.
There is a broad consensus that the LF of galaxies in the field is mainly flat ($\alpha \sim -1$), 
meanwhile a brighter characteristic magnitude $M^{\ast}$ and a steeper faint end slope 
$\alpha$ have been found in galaxy systems for absolute magnitudes in the range
$M_r\la-16$. Other authors argue that a very steep faint end slope, at $M_r\ga-17$, can be measured
in rich clusters and galaxy groups \citep{pop05,gonz05} when using only photometric information.
It should be remarked that the methods that do not use spectroscopic redshifts are sensitive to the background
computation, so, they are less reliable.

%%%%%%%%%%%%%%%%%%%%%%%%%%%%%%%%%%%%%%%%%%%%%%%%%%%%%%%%%%%%%%%%%%%%%%%%%%%%%%%
%%%%%%%%%%%%%%%%%%%%%%%%%%%%%%%%%%%%%%%%%%%%%%%%%%%%%%%%%%%%%%%%%%%%%%%%%%%%%%%
\begin{figure}
\includegraphics[width=90mm]{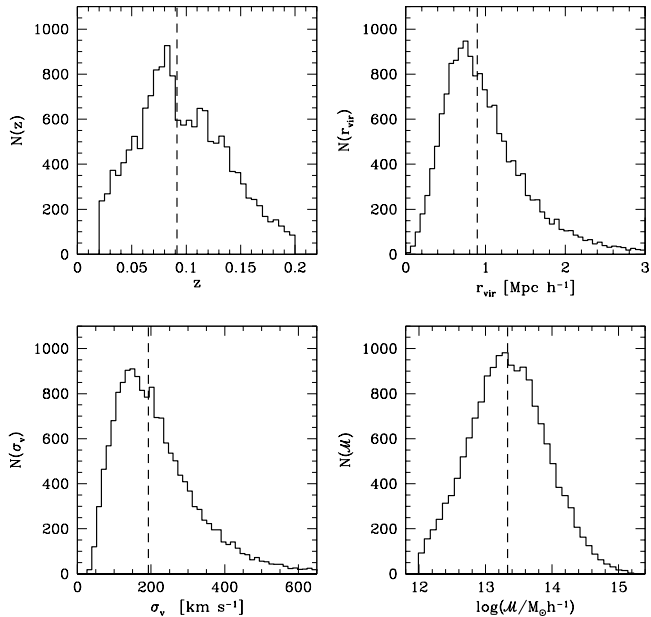}
\caption{
Distributions of the properties SDSS DR7 groups. Redshift distribution ({\em upper left panel}),
virial radii ({\em upper right panel}), line of sight velocity dispersion ({\em lower left panel})
and virial mass ({\em lower right panel}).
Vertical dashed lines are the median of the distributions.
}
\label{fig1}
\end{figure}
%%%%%%%%%%%%%%%%%%%%%%%%%%%%%%%%%%%%%%%%%%%%%%%%%%%%%%%%%%%%%%%%%%%%%%%%%%%%%%%
%%%%%%%%%%%%%%%%%%%%%%%%%%%%%%%%%%%%%%%%%%%%%%%%%%%%%%%%%%%%%%%%%%%%%%%%%%%%%%%

Many works in the last few years have been concentrated on the effect of the 
environment on the LF. For instance, \citet{croton05,hoyle05,xia06,park07,phleps07} studied the
dependence of the LF on the density contrast within spheres of different radii. 
A brightening of the characteristic magnitude and a steepening of the
faint end slope are observed when moving from underdense to overdense regions.
\cite{deng07} found that the dependence of galaxy luminosity on a dense environment is much weaker
than that on an underdense environment. \cite{choi07} argued that the LF shows significant fluctuations
due to large scale structures, while the morphological fraction as a function of luminosity is relatively less
sensitive and thus seems to be more universal. The importance of the large scale environment was also 
established by \cite{yang09} and \cite{tempel09} showing strong environmental dependencies. 

One of the main benefits of working with large spectroscopic samples of galaxies is that they are very suitable
for the construction of large galaxy group catalogues. \citet{zmm06} used the main galaxy sample
of the SDSS DR4 to construct a large galaxy group catalogue, and they have deepen the study of galaxy luminosities as a
function of environment. Their analysis comprised the variation of the Schechter function parameters, for
different galaxy populations, as a function of the galaxy group virial mass. Their results showed
clear variations of $M^{\ast}$ and $\alpha$ with group virial mass, and proved that these variations are mainly 
caused by the red population of galaxies in groups.
More recently, \cite{robot10} studied the dependence of the LF of galaxies in 2dFGRS groups 
on the group virial mass and multiplicity.
They also found clear trends for decreasing $\alpha$ when increasing the masses and/or multiplicity for early type galaxies,
while a much suppressed relation was observed for late type population. 

At present, the largest galaxy redshift survey is the Seventh Data Release of the
Sloan Digital Sky Survey (hereafter DR7; \citealt{dr7}). 
This catalogue covers a very wide area on the sky and has high quality photometric and spectroscopic information.
From this galaxy catalogue we can extract one of the largest galaxy group catalogued to date.
Therefore, the main aim of this work is two fold: firstly, improving the results obtained for the 
mass dependence galaxy LFs in the SDSS DR4 \citep{zmm06} by using the galaxies in groups in the SDSS DR7
in order to obtain more reliable and detailed results; and secondly, analysing galaxy LFs of different galaxy types
using several criteria to classify them as well as study the influence
of local and global environment on the LF. All this information is intended to   
provide a better understanding of galaxies in a wide range of density environments and, consequently, 
to clarify the scenario of galaxy evolution.

The layout of this paper is as follows. In section 2 we describe the
galaxy sample and the group identification process. The detailed analysis of 
the LFs is in section 3. {\bf Finally, in section 4 we discuss possible implications of our results
and summarize them in section 5.}

\section{The sample of galaxies in groups}
The Sloan Digital Sky Survey (SDSS) has validated and made publicly available its
Seventh Data Release (DR7; \citealt{dr7}) which consists of $8423~ {\rm deg}^2$ of 
five-band, $u \ g \ r \ i \ z$, imaging data and 1,374,080 ($8032~ {\rm deg}^2$) spectra of galaxies, 
quasars and stars. The DR7 Main Galaxy Sample (MGS; \citealt{mgs}) 
consists of galaxies with $r-$band Petrosian magnitudes $r\le17.77$ and $r-$band Petrosian half-light surface 
brightnesses $\mu_{50}\le24.5\ {\rm mag\ arcsec^{-2}}$.

Our sample of groups has been identified in the MGS of DR7 following \citet{mz05}.
We performed the identification of groups on a subsample of MGS that includes $\sim650,000$ objects up to a
redshift of $z=0.2$. 
We used a Friends-of-Friends algorithm that links galaxies ($i,j$) which satisfy
$D_{ij}\le D_0 \times R$ and $V_{ij}\le V_0 \times R$, where $D_{ij}$ is their projected distance and
$V_{ij}$ is their line-of-sight velocity difference. The scaling factor $R$ is introduced
in order to take into account the decrement of the galaxy number density due to the
apparent magnitude limit cutoff (e.g. \citealt{hg82,ramella97,merchan02,mz05}). 
We have adopted a transverse linking length $D_0$ 
corresponding to a contour over-density of $\delta \rho/\rho=200$ (this value corresponds to 
an integrated overdensity of $330$ in a $\Lambda$ Cold Dark Matter cosmology), 
a line-of-sight linking length of $V_0=200~\kms$ and a fiducial redshift
of $0.035$. Since it is well known that the MGS is incomplete for $r$-band apparent magnitudes lower
than $14.5$, we have adopted this magnitude as a lower apparent magnitude limit in our algorithm and excluded
galaxies brighter than this from our analyses.

%%%%%%%%%%%%%%%%%%%%%%%%%%%%%%%%%%%%%%%%%%%%%%%%%%%%%%%%%%%%%%%%%%%%%%%%%%%%%%%
%%%%%%%%%%%%%%%%%%%%%%%%%%%%%%%%%%%%%%%%%%%%%%%%%%%%%%%%%%%%%%%%%%%%%%%%%%%%%%%
\begin{figure}
\includegraphics[width=90mm]{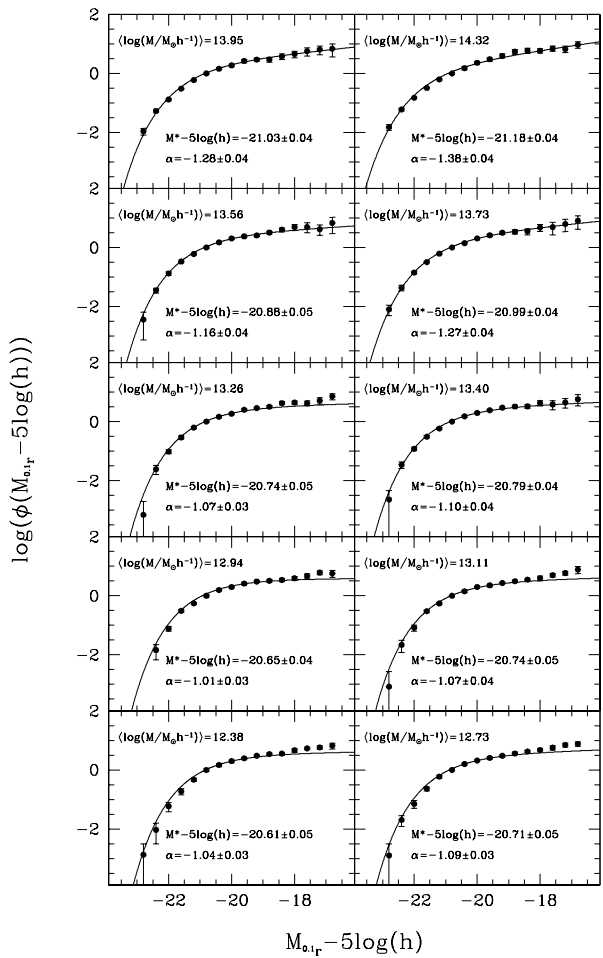}
\caption{
The group mass dependence of the $^{0.1}r-$band LF of galaxies in groups.
Points are $C^-$ estimates of the LF (arbitrary units), while error bars 
are computed using a bootstrap re-sampling technique. Each panel corresponds
to a different group mass bin (labelled by their median mass), and the best 
fitting Schechter function as determined by the STY method.  
}
\label{fig2}
\end{figure}
%%%%%%%%%%%%%%%%%%%%%%%%%%%%%%%%%%%%%%%%%%%%%%%%%%%%%%%%%%%%%%%%%%%%%%%%%%%%%%%
%%%%%%%%%%%%%%%%%%%%%%%%%%%%%%%%%%%%%%%%%%%%%%%%%%%%%%%%%%%%%%%%%%%%%%%%%%%%%%%

As in \citet{mz05}, we have carried out an improvement of the 
rich group identification by performing a second identification (double identification, hereafter $DI$) 
on galaxy groups which have at least ten members. We used a higher density contrast ($\delta \rho/\rho \sim 315$), 
in order to split merged systems or to eliminate spurious member detections (see \citealt{diaz05}). 
The position of the group centres for groups with at least 10 members has been determined by using an iterative 
procedure developed by \citet{diaz05}. 
The procedure defines a new group centre estimation by using the projected local number density 
(using the 5 closest neighbours) of each galaxy member as a weight for their group centric distances, and then 
iterates this estimation by removing galaxies beyond a given distance. The iteration process stops
when the centre location remains unchanged. 

The final group sample comprises 15,961 groups with at least 4 members,
adding up to 103,342 galaxies. 
Following \citet{merchan02}, group virial masses were computed as
${\cal M}=\sigma^2R_{\rm vir}/G$,
where $R_{\rm vir}$ is the virial radius of the system, and $\sigma$ is the velocity dispersion of
member galaxies \citep{limber60}.  
The velocity dispersion $\sigma$ is estimated using the line-of-sight velocity dispersion $\sigma_{v}$,
$\sigma=\sqrt{3}\sigma_v$. To compute $\sigma_v$ we use the methods described by \citet{beers90}.
The group sample has a median redshift, line-of-sight velocity dispersion, virial 
mass and virial radius of 0.09, 
$193 \ \kms$, $2.1\times10^{13} \ h^{-1} \ {\cal M}_{\odot}$, and $0.9 \ \mpc$, respectively. Figure~\ref{fig1} 
shows the distributions of these group properties for our whole sample.

Galaxy magnitudes used throughout this work are Petrosian magnitudes \citep{petrosian},
and have been corrected for Galactic extinction using the maps by \citet{sch98}. Absolute magnitudes have been 
computed assuming a flat cosmological model with parameters $\Omega_0=0.3$, 
$\Omega_{\Lambda}=0.7$ and $H_0=100~h~{\rm km~s^{-1}~Mpc^{-1}}$ 
and $K-$corrected using the method of  \citet{blantonk}~({\small KCORRECT} 
version 4.1). We have adopted a band shift to a redshift $0.1$, i.e. to approximately the 
mean redshift of the MGS of SDSS. 
We have also included evolution corrections to this redshift following \citet{blantonlf}.
Throughout this work we will refer to these shifted bands 
as $^{0.1}u, \ ^{0.1}g, \ ^{0.1}r, \ ^{0.1}i$, and $^{0.1}z$. All magnitudes are in the AB system.

%%%%%%%%%%%%%%%%%%%%%%%%%%%%%%%%%%%%%%%%%%%%%%%%%%%%%%%%%%%%%%%%%%%%%%%%%%%%%%%
%%%%%%%%%%%%%%%%%%%%%%%%%%%%%%%%%%%%%%%%%%%%%%%%%%%%%%%%%%%%%%%%%%%%%%%%%%%%%%%
\begin{figure*}
\includegraphics[width=150mm]{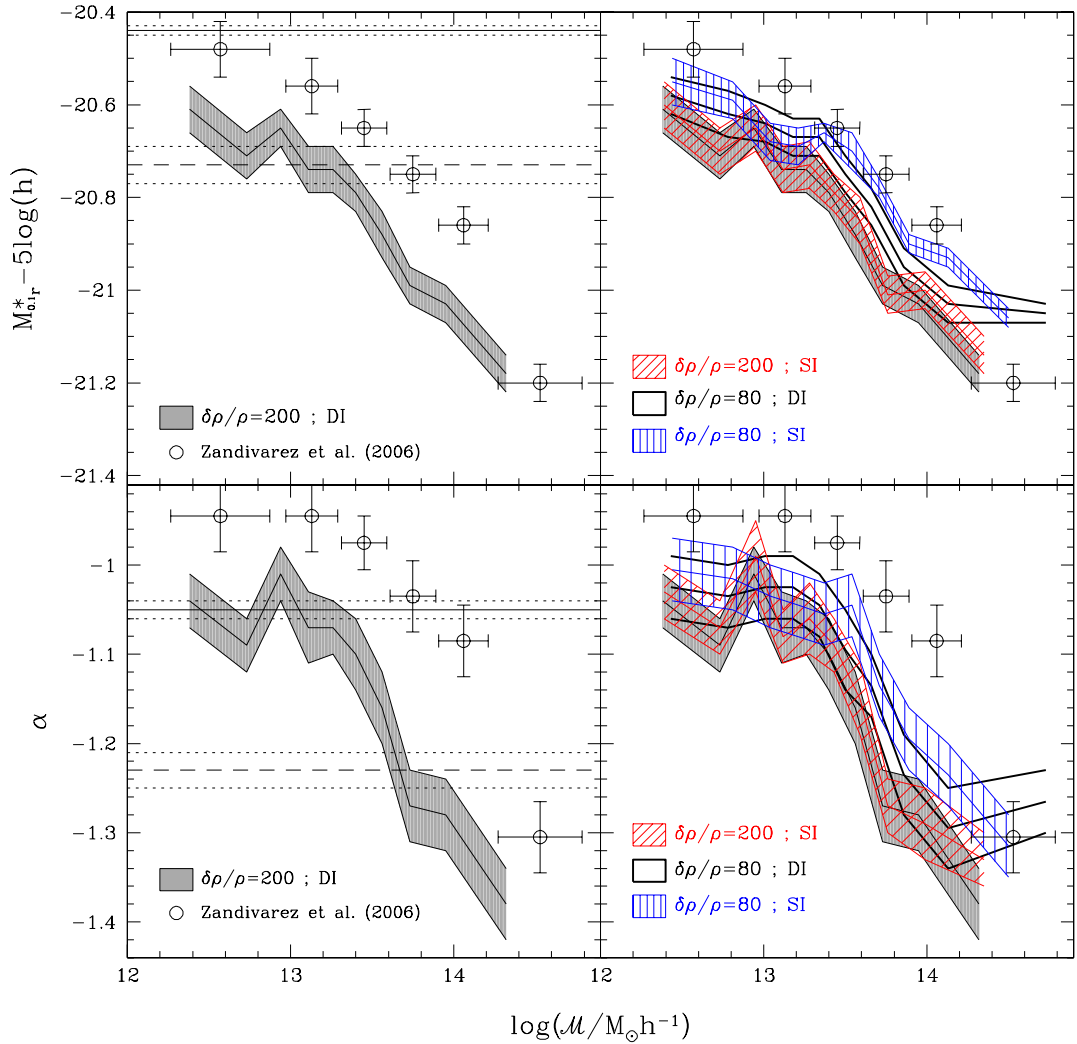}
\caption{{\em Left panels:} STY best fitting Schechter functions parameters in the $^{0.1}r$ band as 
a function of group mass for our sample of SDSS DR7 groups. Error bars are the projection of 
1$\sigma$ joint error ellipse onto the $\alpha$ and $M^{\ast}$ axes. {\em Horizontal lines} are the field LF parameters
obtained by \citet{blantonlf} ({\em solid lines}) and \citet{montero09} ({\em dashed lines}) along with their
$1\sigma$ standard deviation ({\em dotted lines}). {\em Open circles} are the results by \citet{zmm06}. 
{\em Right panels:} the dependence of
the Schechter function parameters on different group identification procedures, see text 
for details. 
}
\label{fig3}
\end{figure*}
%%%%%%%%%%%%%%%%%%%%%%%%%%%%%%%%%%%%%%%%%%%%%%%%%%%%%%%%%%%%%%%%%%%%%%%%%%%%%%%
%%%%%%%%%%%%%%%%%%%%%%%%%%%%%%%%%%%%%%%%%%%%%%%%%%%%%%%%%%%%%%%%%%%%%%%%%%%%%%%

\section{The group mass dependence of the luminosity function of galaxies in groups}
In this section we study the luminosity function of
galaxies in our sample of SDSS DR7 groups as a function of group virial mass and its 
dependence on: group identification procedure, photometric band, galaxy type,
galaxy position within the groups and the large scale environment surrounding the
groups.

For computing the LFs we adopt two different methods: to visualise the LF, we use the 
non parametric $C^-$ method \citep{lb71,cho87}, which is the most robust estimator
being less affected by different values of the faint end slope and sample size
\citep{w97}; and the STY method \citep{sty79} which is reliable for fitting analytic
functions without binning the data. In all cases studied in this paper, we find that the
Schechter parametrisation of the LF \citep{schechter76} is appropriate, therefore, most of our
findings below are expressed only in terms of the values of the Schechter function shape parameters 
$\alpha$ and $M^{\ast}$. 

\subsection{The overall group galaxy population LF}
We have computed the $^{0.1}r-$band LFs of galaxies in groups for the full sample of groups in the SDSS DR7
as a function of group virial mass. The corresponding LFs are shown in Fig.~\ref{fig2},
where each panel corresponds to a different group virial mass bin from lower ({\em left bottom panel}) to 
higher ({\em right top panel}) mass values. The solid lines show the STY best fit Schechter LFs, and
the corresponding STY $\alpha$ and $M^{\ast}$ values are quoted therein.
A clearer way to analyse the variation of the LF with mass is to represent the best fitting parameters 
($\alpha$ and $M^{\ast}$) as a function of virial mass. 
In the {\em left panels} of Fig.~\ref{fig3} we show the $^{0.1}r-$band LF's parameters of galaxies in groups.
Our results show a clear brightening in the characteristic magnitude 
and  a decreasing faint end slope with mass. The amplitude variations ($\Delta M^{\ast} \sim 0.6$
and $\Delta \alpha \sim 0.4$) are fully consistent with the results obtained by \citet{zmm06}, although
the $M^{\ast}$ and $\alpha$ parameters are slightly shifted ($\sim 0.1$) to brighter and steeper 
values, respectively. 
These differences are examined in detail in the following section.
Even when these trends are in agreement with the results obtained by 
\citet{yang05}, \citet{zmm06} and \citet{robot10}, the large number of groups used in our analysis 
allow us to obtain quite more clear and detailed determinations than those obtained in previous works.  

%%%%%%%%%%%%%%%%%%%%%%%%%%%%%%%%%%%%%%%%%%%%%%%%%%%%%%%%%%%%%%%%%%%%%%%%%%%%%%%
%%%%%%%%%%%%%%%%%%%%%%%%%%%%%%%%%%%%%%%%%%%%%%%%%%%%%%%%%%%%%%%%%%%%%%%%%%%%%%%
\begin{figure}
\includegraphics[width=90mm]{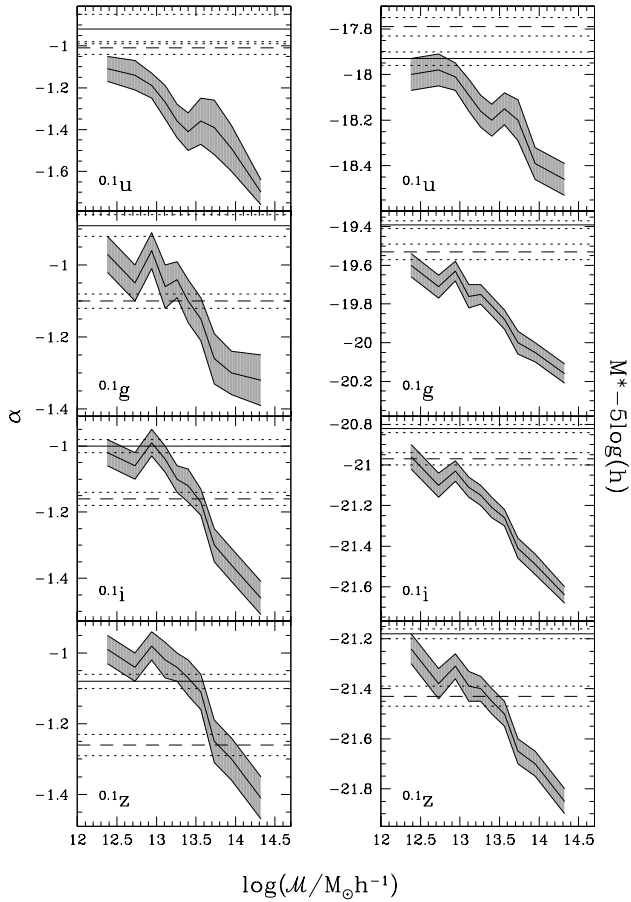}
\caption{STY best fitting Schechter functions parameters in the $^{0.1}u$, $^{0.1}g$, $^{0.1}i$ and $^{0.1}z$ 
bands as a function of group mass for our sample of SDSS DR7 groups. 
Error bars are the projection of 1$\sigma$ joint error ellipse onto the $\alpha$ and $M^{\ast}$ axes.
{\em Horizontal lines} are the field LF parameters obtained by \citet{blantonlf} ({\em solid lines}) and 
\citet{montero09} ({\em dashed lines}) along with their $1\sigma$ standard deviation ({\em dotted lines}).}
\label{fig4}
\end{figure}
%%%%%%%%%%%%%%%%%%%%%%%%%%%%%%%%%%%%%%%%%%%%%%%%%%%%%%%%%%%%%%%%%%%%%%%%%%%%%%%
%%%%%%%%%%%%%%%%%%%%%%%%%%%%%%%%%%%%%%%%%%%%%%%%%%%%%%%%%%%%%%%%%%%%%%%%%%%%%%%

\subsubsection{The dependence of the LF on the group identification procedure}
As described in Sect. 2, we have chosen to identify galaxy groups with a contour density
contrast $\delta\rho/\rho=200$. This particular value differs from the one adopted in our 
previous work \citep{zmm06} where groups with $\delta\rho/\rho=80$ were used. The choice
adopted in the current work relies on the idea of considering an even more reliable sample of 
galaxy groups to perform the luminosity statistics. Since the development of the Friends-of-Friends 
algorithm to identify groups in redshift space, the main reason to choose
a value of $\delta\rho/\rho=80$ was to include as many loose systems as possible in order to
obtain larger samples, and without changing the main physical properties of galaxy groups 
(see for instance \citealt{ramella97}). 
In a previous work, \citet{merchan02} have shown that the larger the $\delta\rho/\rho$
used in the identification process, the higher the reliability of identifying real systems. 
In that work they have chosen a $\delta\rho/\rho=80$ just to include as many systems as possible
at the expense of some reliability in the groups identified.
Using $\delta\rho/\rho=80$ basically selects galaxies within a radius that corresponds to the geometric mean between
the virial and the turnaround radii, therefore, galaxy groups include 
infalling galaxies that can bias the group definition, mass estimate and properties \citep{mamon07}.
Our purpose in this work is to avoid the loosest systems which are the least reliable in both, their 
probability of being real systems and the stability of their physical properties, and also exclude infalling
galaxies, thus reducing biases in the luminosity statistics.

It is interesting to investigate whether the choice of $\delta\rho/\rho$ or the refinement procedure
in the identification could introduce observable effects on the LFs.  
Hence, we have performed different group identifications varying $\delta\rho/\rho$, and 
applying or not the refinement identification process (DI) described in Sect.~2. The contour
overdensities adopted are 80 and 200 while performing or not the refinement is named as double
identification (DI) or single identification (SI) respectively. The results obtained with
these new group samples are shown in the {\em right panels} of Fig.~\ref{fig3}.
Analysing the trends observed for the characteristic magnitude ({\em upper panel}), it can be seen that the 
group identification with $\delta\rho/\rho=80$ produces groups with typically fainter 
$M^{\ast}$ than those observed for $\delta\rho/\rho=200$, and such difference is more
pronounced when just a single identification is performed. For the faint end slope trends
({\em lower panel}), the behaviour is similar but less notorious than the described above.
The observed differences in both Schechter parameters, arise primarily from using a lower density contrast,
which has two main effects: the presence of not virialised
systems whose existence and physical properties are less reliable, and the inclusion of galaxies in the 
outer parts of groups. The presence of loose systems can affect the LF over the whole range of masses,
while the extended outer parts of groups should have a stronger effect in $M^{\ast}$ for high mass groups.
Evidence of the latter is provided in our analysis of the variation of the LF as a function of group centric distance 
in Sect. 3.3 below.  The low density contrast results are in a better agreement with the previous
results obtained by \citet{zmm06}. The remaining differences, mainly observed in $\alpha$, could be attributed
to an increment of $\sim75\%$ in the number of galaxies between DR4 and DR7.
For the purposes of this work we prefer to use $\delta\rho/\rho=200$ in our analyses in order 
to minimise possible biases.

\subsubsection{The LF in all SDSS photometric bands} 
A full photometrical description of galaxies in groups is achieved by studying the LFs 
for different photometric bands as a function of group mass. Being the SDSS galaxy sample
a $r-$band magnitude limited one, it does not mean that such sample is magnitude
limited in the other SDSS bands. Therefore, we have introduced apparent magnitude cuts-offs
in each SDSS photometric band in order to build complete magnitude limited samples
of galaxies. The corresponding cut-offs in each band (see table in Appendix A)
are obtained from the analysis of the galaxy number counts in each SDSS DR7 band.
The adopted limits are more conservative than those adopted by \cite{montero09} for the SDSS 
DR6 (see their Table 2). Since $K-$corrections for each band are only reliable within certain
redshift ranges, we have also adopted the redshift cut-offs for each band introduced by 
\cite{blantonlf}, which are consistent with those used by \cite{montero09}. 

The resulting Schechter parameters as a function of groups masses for each photometric SDSS band
are shown in Fig.~\ref{fig4}. We observe a similar behaviour as the obtained in the $r-$band for 
both Schechter parameters as a function of the group masses. The characteristic magnitude shows ({\em left panels}) 
approximately the same increment from low to high group masses in the $\gb$, $\ib$ and
$\zb$ bands ($\sim 0.6$ mag), similar to the $\rb$ results. 
This brightening is smaller ($\sim 0.5$) in the $\ub$ band.
The largest variation of the faint end slope as a function of mass is observed in the $u-$band 
($\sim 0.6$) while the other bands show roughly the same increment ($\sim 0.4$) and also similar to the $\rb$ band.

As a comparison with field LF, we show in Figs. \ref{fig3} and \ref{fig4} the results by \cite{blantonlf} 
and \cite{montero09} for galaxies in the SDSS DR1 and DR6, respectively.
Since those galaxy samples comprise all MGS galaxies in each data release, 
the results obtained for the Schechter parameters are expected to be representative of the low mass groups tail of our analyses. 
This is clearly observed for both Schechter parameters when analysing the \cite{blantonlf} results, while
this is true only for the characteristic magnitudes obtained by \cite{montero09}. The faint end slopes obtained
by \cite{montero09} are typically steeper than ours. However, analysing their LFs
and comparing them with those obtained by \cite{blantonlf}, we observe that
the behaviours of both determinations, at the faint end slope, are quite similar (see Figs. 
7 and 8 of \citealt{montero09}). The differences obtained in the fitted faint end slope are the result of a small 
brightening in the characteristic magnitude in the \cite{montero09} compared to \cite{blantonlf}.

%%%%%%%%%%%%%%%%%%%%%%%%%%%%%%%%%%%%%%%%%%%%%%%%%%%%%%%%%%%%%%%%%%%%%%%%%%%%%%%
%%%%%%%%%%%%%%%%%%%%%%%%%%%%%%%%%%%%%%%%%%%%%%%%%%%%%%%%%%%%%%%%%%%%%%%%%%%%%%%
\begin{figure}
\includegraphics[width=90mm]{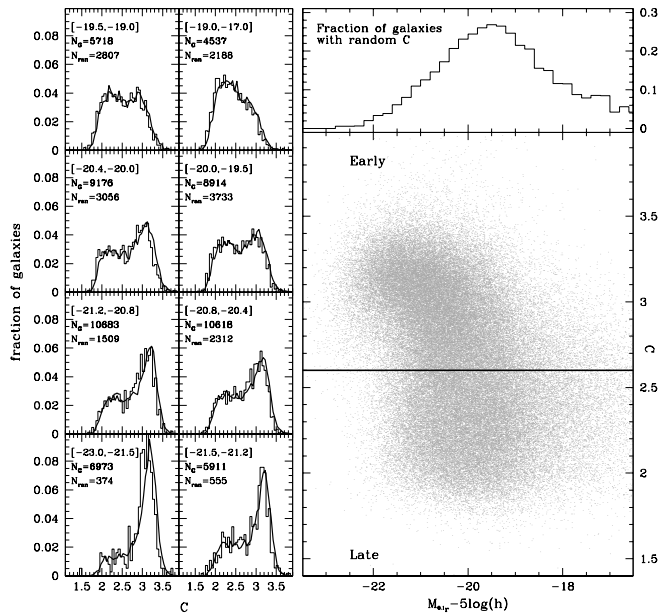}
\caption{
Classifying galaxies into early and late types according to the concentration parameter
$C$. {\em Left panels} show the distributions of $C$ for MGS galaxies with apparent 
magnitude $r\le 16$ in different absolute magnitude bins ({\em thick solid lines}). 
These distributions are used to randomly assign $C$ values to group galaxies whose $r_{50}$ values
are strongly affected by seeing (see text for details). The resulting distributions of randomly assigned
$C$ are shown as {\em thin line} histograms. 
{\em Right upper panel} shows the fraction of galaxies in groups with randomly assigned $C$ as
a function of absolute magnitude.
{\em Right lower panel} shows $C$ vs absolute magnitude
for all galaxies in groups. About $16\%$ of these galaxies have random $C$ values. 
The {\em horizontal line} is $C=2.6$ that splits galaxies into early and late types.
}
\label{fig5}
\end{figure}
%%%%%%%%%%%%%%%%%%%%%%%%%%%%%%%%%%%%%%%%%%%%%%%%%%%%%%%%%%%%%%%%%%%%%%%%%%%%%%%
%%%%%%%%%%%%%%%%%%%%%%%%%%%%%%%%%%%%%%%%%%%%%%%%%%%%%%%%%%%%%%%%%%%%%%%%%%%%%%%
\begin{figure}
\includegraphics[width=90mm]{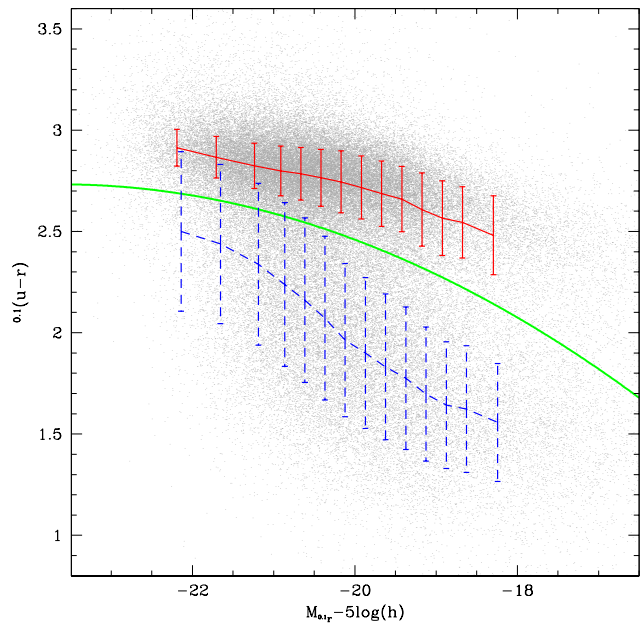}
\caption{The colour-magnitude diagram of galaxies in SDSS DR7 groups. 
{\em Solid red lines} are the centres of the Gaussian functions that represent the
colour distribution of the SDSS DR7 MGS overall red galaxy sequence as a 
function of absolute magnitude.
{\em Dashed blue lines} are the corresponding values for the blue galaxy sequence.
In both cases we show the width of the Gaussian functions as error bars. 
The {\em green thick solid line} is the function we use to split galaxies into Red and
Blue subsamples (see text for details).
}
\label{fig6}
\end{figure}
%%%%%%%%%%%%%%%%%%%%%%%%%%%%%%%%%%%%%%%%%%%%%%%%%%%%%%%%%%%%%%%%%%%%%%%%%%%%%%%
%%%%%%%%%%%%%%%%%%%%%%%%%%%%%%%%%%%%%%%%%%%%%%%%%%%%%%%%%%%%%%%%%%%%%%%%%%%%%%%

\subsection{The luminosity function of different galaxy types}
The galaxy population in the local Universe can be broadly described in 
terms of two different classes of objects, early and late types, which can be distinguished
by their morphology, star formation rate and colour (e.g. \citealt{strat01,balogh04,baldry04}).
Evidence has been found that the properties of early-type galaxies are almost independent
of the environment (e.g. \citealt{dress97,bernardi03,mcm10}). This has also been reported
for late types (e.g. \citealt{biviano90,zmm06}).
There is conclusive evidence that the bi-modality in galaxy colours is already in place at $z\sim 1$,
and that the fractions of early and late types are different compared to the local universe.
\citep{bell04,tanaka05}. A number of physical mechanisms related to environment 
can be responsible for this bimodality by transforming galaxies from late to early types
(e.g., ram pressure, galaxy harassment, interactions with potential wells)
or by truncating their star formation (e.g. strangulation).

In this section, we explore the mass dependence of the LF of different
galaxy types. Galaxies are classified into different types according to their concentration index and colour, with the 
aim of shedding light on the effect of group environment in the transformation of galaxies
and the build up of the bimodality observed in the galaxy population.

\subsubsection{Early and Late galaxies according to the concentration index $C$}
The concentration index is defined as the ratio of the radii enclosing
90 and 50 percent of the Petrosian flux, $C=r_{90}/r_{50}$. This parameter is related
to the light distribution within a galaxy. Typically, early-type galaxies
have $C>2.6$, while for late-types $C<2.6$ \citep{strat01}, thus, $C$ is an indicator of morphology.
We split group galaxies into early and late types according to their $C$ parameter,
and compute the group mass dependence of the LF of galaxies of each type.

Special care must be taken when using $C$ to classify galaxies. 
The effects of seeing in the measurement of $r_{50}$ and $r_{90}$, and thus in $C$,
have to be considered for galaxies with relatively small angular sizes. 
The average seeing in the SDSS is below a conservative value of $1.6\arcsec$, thus for
galaxies that have $r_{50}\la 1.6\arcsec$, $C$ can be unreliable.
We have found that $\sim 16\%$ of all MGS galaxies with redshifts $z\le0.2$ (our 
redshift upper limit) have $r_{50}$ values below $1.6\arcsec$.
This number drops to less than $0.7\%$ if we impose a further apparent magnitude cut-off
of $r\le 16$.
We have used volume limited subsamples of these brighter MGS galaxies to compute the distribution of
$C$ for different luminosity bins (see {\em left panels} of Fig.~\ref{fig5}). 
These distributions are the used to randomly assign $C$ values to those galaxies that had unreliable
values of $C$ ($r_{50}\le 1.6\arcsec$). {\em Right upper panel} of Fig.~\ref{fig5} shows the fraction of galaxies
in group with randomly assigned $C$ values as a function of absolute magnitude.
We used Kolmogorov-Smirnov test to choose the $C$ random assignations that reproduce the best the
observed $C$ distributions per bin of absolute magnitude. These distributions are shown
by histograms in the {\rm left panels} of Fig.~\ref{fig5}. Finally, {\em right lower panel}
shows the scatter plot of $C$ vs. absolute magnitude for all the galaxies in groups
in SDSS DR7, i.e., those whose $C$ was assigned randomly and those whose $C$ was reliable
from the very beginning.

\begin{figure*}
\includegraphics[width=180mm]{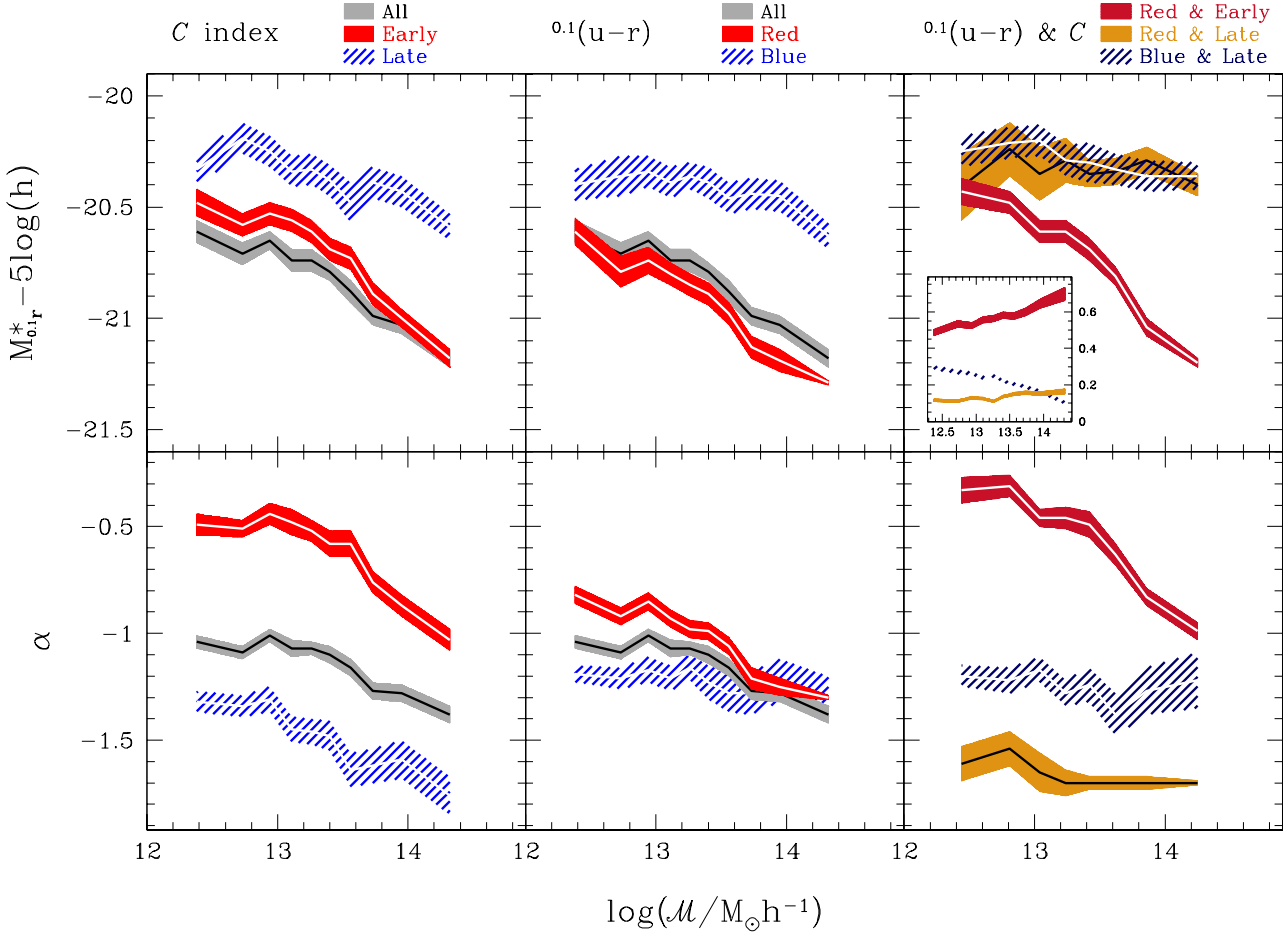}
\caption{STY best fitting Schechter function parameters in the $^{0.1}r$ band as a function 
of group mass for different galaxy populations. 
{\em Left panels} show the results obtained from classifying galaxies into early
and late types according to their concentration index $C$.
In the {\em middle panels} galaxies are classified as belonging to the red or blue
sequences according to their $^{0.1}(u-r)$ colour. 
{\em Right panels} show the results of classifying galaxies using both $C$ and colour.
Error bars are the projection of 1$\sigma$ joint error ellipse onto the $\alpha$ and $M^{\ast}$ axes. 
{\em Inset panel} shows the fractions of different types of $M_{^{0.1}r}-5\log(h)\le -19.5$ and $z\le0.09$
galaxies as a function of group mass.
}
\label{fig7}
\end{figure*}

Once all the galaxies in groups have $C$ values assigned, we classify them into early or late types
(see {\em right panel} of Fig.~\ref{fig5}) and compute the mass dependence of the $^{0.1}r-$band LF of each galaxy 
type. We show in the {\em left panels} of Fig.~\ref{fig7} the group mass dependence of the 
parameters $\alpha$ and $M^{\ast}$.
We also show, for comparison, the results from the overall population of galaxies in groups
determined in the previous section. Clearly, early and late type galaxies have different
LFs over the whole group mass range. Regarding the characteristic magnitude, $M^{\ast}$,
it is systematically brighter for early types, the difference ranging from $\sim0.2$ to
$\sim0.6$ mag. Also, the dependence of $M^{\ast}$ on mass is stronger for early types. This behaviour
closely follows the trend of the overall galaxy population, despite being fainter
over most of the mass range.
The faint end slope, $\alpha$, of early and late types differ in $\sim0.8$ over the
whole mass range, being, as expected, larger in absolute value for late types.  Their trends
as a function of group mass are roughly parallel to each other. 

%%%%%%%%%%%%%%%%%%%%%%%%%%%%%%%%%%%%%%%%%%%%%%%%%%%%%%%%%%%%%%%%%%%%%%%%%%%%%%%
%%%%%%%%%%%%%%%%%%%%%%%%%%%%%%%%%%%%%%%%%%%%%%%%%%%%%%%%%%%%%%%%%%%%%%%%%%%%%%%
%%%%%%%%%%%%%%%%%%%%%%%%%%%%%%%%%%%%%%%%%%%%%%%%%%%%%%%%%%%%%%%%%%%%%%%%%%%%%%%
%%%%%%%%%%%%%%%%%%%%%%%%%%%%%%%%%%%%%%%%%%%%%%%%%%%%%%%%%%%%%%%%%%%%%%%%%%%%%%%

\subsubsection{Red and Blue galaxies according to the $^{0.1}(u-r)$ colour}
At a fixed luminosity, the colour distribution of galaxies can be remarkably
well described as the combination of two Gaussian functions that represent the contributions
from galaxies in the blue and the red sequence (e.g. \citealt{baldry04,balogh04}).
We use this bi-modality of galaxy colours to split galaxies into red and blue
populations. 

We proceed as follows: firstly, we fit the two-Gaussian model to the
 $^{0.1}(u-r)$ colour distribution of the MGS galaxies in different absolute magnitude bins.
For this purpose, we include in each absolute magnitude bin the largest volume limited
subsample of MGS galaxies.
The fitting procedure gives, for a given luminosity bin, 6 parameters that measure the amplitude,
mean value and width of the two Gaussian functions. In Fig.~\ref{fig6} we
show the mean colour of the blue and red populations as a function of 
absolute magnitude and the corresponding half-maximum widths as error-bars,
superimposed over the colour-magnitude diagram of MGS galaxies.
Secondly, we looked for the colour that separates between blue 
and red populations as a function of absolute magnitude: for each luminosity bin, we
compute the colour value at which the two Gaussian functions intersect, i.e., 
the colour value that gives a galaxy the same probability of belonging to
the red or to the blue population. We have found that these 'equal-probability' points 
can be very well fitted by a second degree polynomial: $P(x)=-0.02x^2-0.15x+2.46$, where
$x=M_{^{0.1}r}-5\log(h)+20$, shown as a {\em thick green line}
in Fig.~\ref{fig6}. Finally, we use this polynomial to classify galaxies as red or blue
for the computation of the  $^{0.1}r-$band LF. 

In the {\em middle panels} of Fig.~\ref{fig7} we show the resulting parameters  
$\alpha$ and $M^{\ast}$ as a function of group mass for galaxies classified as blue
and red and the comparison with the results for the overall galaxy population of groups (Sect. 3.1).
The first conclusion that can be extracted from the comparison of these panels with the left panels
is that, classifying galaxies according to colour is not as closely related to classifying galaxies according
to concentration index as one may, a priori, think, since the results for blue/red differ 
from those of early/late types discussed above.
The mass dependence of $M^{\ast}$ and $\alpha$ for the red and blue sequences qualitatively
agree with our previous findings using DR4 groups \citep{zmm06}. 
The red sequence $M^{\ast}$ is brighter than its blue counterpart over the
whole mass range in $\sim0.2-0.8$ mag, and has a stronger dependence on mass.
It is also brighter than that of the whole sample, at least for masses 
$\ga 1.6\times10^{13}h^{-1}M_{\odot}$.
The blue sequence $M^{\ast}$ shows a weak dependence on mass.
Regarding $\alpha$, the results for the blue sequence are independent of group mass,
showing a constant value of $\sim -1.2$, whereas the red sequence
shows a variation of $\sim0.5$ over the whole mass range.

The fact that the blue sequence LF shows such a weak dependence on the mass
of the systems supports the existing idea of a blue population whose properties
are basically independent of the environment.

\subsubsection{The role of red spheroids}

Splitting galaxies into two populations according to a morphological or a colour criterion has
led to remarkably different faint end slopes (Fig.~\ref{fig7}). This result indicates that
the correspondence between galaxy colour and morphology
is not perfect, as was also pointed out by \citet{bundy10}. In that work, the authors studied a population
of red sequence galaxies with disk-like morphologies taken from the COSMOS survey. They found that
up to half of the red sequence galaxies harbour disks in the redshift range $0.3\le z\le 1.2$, 
and that their contribution to the red sequence declines with time.

In the {\em right panels} of Fig.~\ref{fig7}, we show the group mass dependence of the LF
for galaxies classified according to both: $C$ and colour, simultaneously.
We show $\alpha$ and $M^{\ast}$ as a function of group mass
for red-early galaxies, red-late galaxies (i.e. passive disks) and blue-late galaxies. 
We did not introduce in this figure the results for galaxies classified as blue-early because 
there are only few of them which turns LF computation very noisy. 
It is clear from these panels that for both late types, blue-late and red-late, the LF parameters 
show no clear group mass dependence. Both types have similar $M^{\ast}$, and passive disks have 
an LF steeper ($\alpha\sim-1.6$) than that of star forming disks ($\alpha\sim -1.2$). 
The only type that displays strong mass dependence of its LF is the red-early type, dominated by red spheroids.

%%%%%%%%%%%%%%%%%%%%%%%%%%%%%%%%%%%%%%%%%%%%%%%%%%%%%%%%%%%%%%%%%%%%%%%%%%%%%%%
%%%%%%%%%%%%%%%%%%%%%%%%%%%%%%%%%%%%%%%%%%%%%%%%%%%%%%%%%%%%%%%%%%%%%%%%%%%%%%%
\begin{figure}
\includegraphics[width=90mm]{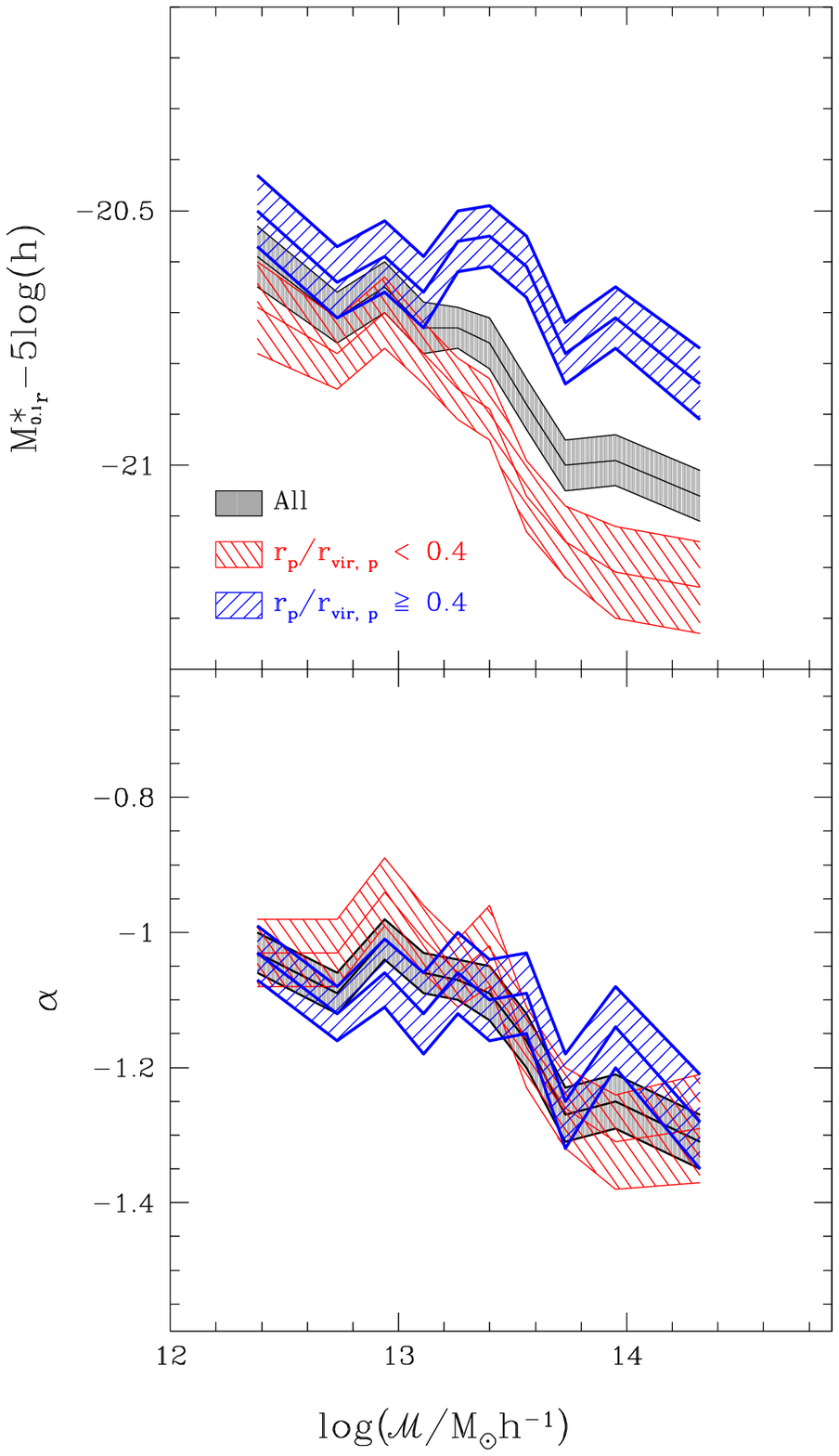}
\caption{
STY best fitting Schechter function parameters in the $^{0.1}r$ band as a function of group mass
for galaxies in the inner and outer regions of groups according to their normalised 
projected group centric distance. Error bars are the projection of 1$\sigma$ joint error ellipse onto each axis.
}
\label{fig8}
\end{figure}
%%%%%%%%%%%%%%%%%%%%%%%%%%%%%%%%%%%%%%%%%%%%%%%%%%%%%%%%%%%%%%%%%%%%%%%%%%%%%%%
%%%%%%%%%%%%%%%%%%%%%%%%%%%%%%%%%%%%%%%%%%%%%%%%%%%%%%%%%%%%%%%%%%%%%%%%%%%%%%%

The general trend of the LF as a function of mass and the uneven behaviours of the LFs of galaxies 
classified according to morphology or 
colour can be understood in light of these results and the {\em inset panel} in Fig.~\ref{fig7},
in which we show the fraction of the 3 colour-morphology types as a function of group mass.
Red-early galaxies have a LF that strongly depends on mass, its $M^{\ast}$ brightens in $\sim 0.8$ mag
and its $\alpha$ steepens in $\sim 0.7$ over the mass range we probe, and also, the fraction of these galaxies 
increases with mass. This galaxies only can account for the trends observed in the overall group LF. 
When classifying galaxies according to $C$, early type galaxies are dominated in number by the
red-early type, therefore it explains the similarity between the trends seen in {\em red} in the {\em left panels} and 
in {\em dark red} in {\em right panels} of Fig.~\ref{fig7}. On the other hand, late type galaxies are a 
mix of blue-late and red-late types, the former dominating at low mass and the latter
increasing their contribution with mass. 
The steepening of the LF of late type galaxies with mass is due to both:
the decreasing fraction of blue-late galaxies and the increasing fraction of passive disks with mass.
As regard to the colour classification, the lack of mass dependence of $\alpha$ of blue galaxies is 
mirroring the fact that they are mostly blue-late types. On the other hand, since the red sequence
includes passive disks, its $\alpha$ is systematically larger in absolute value than that of late
galaxies.

\subsection{The luminosity function dependence on group centric distance}

Galaxies in the inner parts of clusters of galaxies are known to be statistically different
in their physical properties from those in the outer parts (e.g. \citealt{whit91,whit93,dom01}). 
This has also been found in less massive groups of galaxies (e.g. \citealt{dominguez02}). 
In order to deepen our understanding on the galaxy segregation within groups, in this part 
of the paper we explore the dependence of the LF on the position of galaxies inside the groups.

For each galaxy in our group sample we compute its projected distance to the
group centre in units of the group virial radius. We find that the median of the
distribution of these normalised projected distances is 0.4. We use this value
to split galaxies in groups into two classes: galaxies in the inner and outer 
parts of groups, and compute the group mass dependence of the $^{0.1}r-$band LF for each class.
The resulting STY best fit $M^{\ast}$ and $\alpha$ parameters as a function of mass
are shown in Fig.~\ref{fig8}. Given the estimated error-bars, the only significant
difference in the LF of galaxies in the inner and outer parts of groups is seen in
$M^{\ast}$ for groups more massive than $\sim 10^{13}h^{-1}M_{\odot}$. Galaxies in the inner
parts of groups have a brighter $M^{\ast}$, as expected. This difference tends to increase 
with mass since galaxies in the inner parts show a strong dependence on mass while the outers do not. 
On the other hand, there is no significant differences in the behaviour of the faint end slope for inner
and outer group galaxy members.

\subsection{The luminosity function dependence on large scale environment}

So far, all our analyses were carried out focusing on the influence of the
intra-group medium in the Schechter parameters and their variation as a function of the 
group mass. The final issue that arises is whether the influence of the large scale environment
could play a role in the Schechter parameters for different groups. 

In order to quantify this effect, we split the sample of groups into two: groups with
high and low dense large scale environment, where the density of the environment is characterised by a number
density defined as follows. We compute the number density of galaxies in a cylindrical volume around 
groups, $\rho_f$, where 
$f$ is a multiplicative factor used to compute the volume. On the sky plane, the cylinder is defined by the
circle with a radius $f$ times the virial radius of the group measured from its centre
($R=fr_{vir}$), while on the line-of-sight, the cylinder has length defined by $f$ times the velocity dispersion, 
above and below the mean redshift of the group ($L=2f\sigma_v$). 
Due to the irregular shape of the galaxy catalogue, the solid angle of this cylinder is
estimated by taking into account the angular mask of the SDSS DR7. It should be noted that $\rho_f$
is computed subtracting the number density volume of the group, computed by using $f=1$.
Since the sample of galaxies in the catalogue is 
magnitude limited, in order to obtain proper estimations of the number densities we have defined 
a volume limited sample of galaxies to compute densities, using only those with $z<0.11$ and
$M_{^{0.1}r}-5\log(h)<-20$. We have also introduced a lower redshift cutoff ($z>0.04$) since
the lower magnitude cut-off of 14.5 eliminates almost all galaxies in that region. Finally, the sample
of groups is also restricted in the redshift range in order to have properly defined cylindrical volumes,
i.e., we have used only those groups with mean redshift in the interval 
$\left[ 0.04+f\sigma_v/c \ ; \ 0.11-f\sigma_v/c \right]$, where $c$ is the speed of light.

\begin{figure*}
\includegraphics[width=180mm]{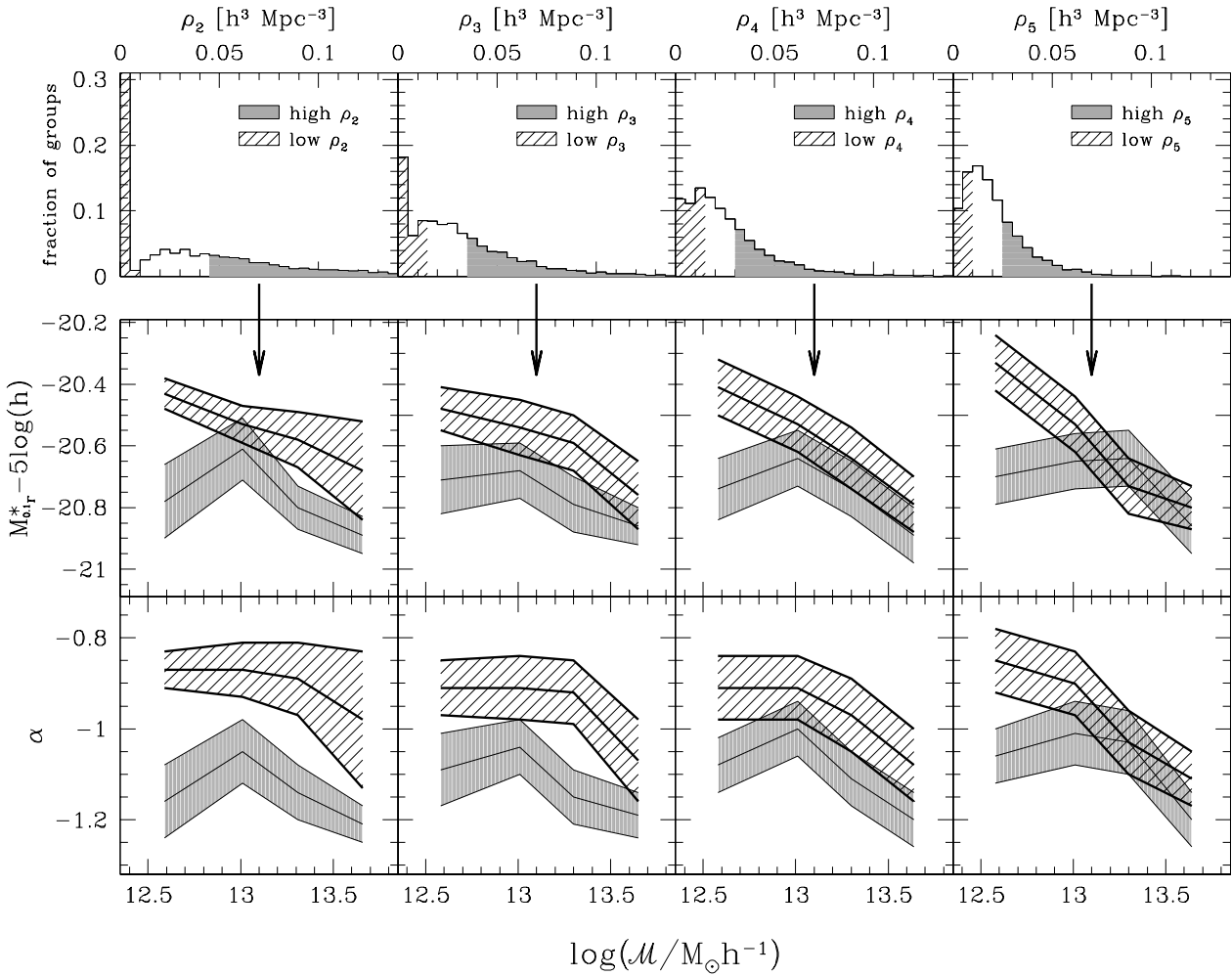}
\caption{The environmental dependence of the galaxy LF.
{\em Upper panels:} distributions of environmental density of groups, $\rho_f$. Densities are computed as
the number of galaxies brighter than $M_{^{0.1}r}-5\log(h)=-20$ within cylindrical volumes defined by $f$ times
the virial radius and the line-of-sight velocity dispersion of groups (see text for details).
From left to right we move from smaller to larger volumes.
Shaded (striped) areas are the subsamples of groups in high (low) density environments.  
{\em Middle and lower panels:} STY best fit Schechter function parameters in the $^{0.1}r$ band as 
a function of group mass for galaxies in the corresponding low and high density group subsamples. 
}
\label{fig9}
\end{figure*}

We have used four different values $f \ = \ 2, \ 3, \ 4$ and $5$. After applying the
corresponding restrictions, each sample comprises $\sim 8500$ groups which have been 
split in low ($\rho_f$ below the $33^{th}$ percentile) and high ($\rho_f$ above the $66^{th}$ 
percentile) large scale density subsamples of groups. The distribution of each type of density and their
corresponding subsamples are shown in the {\em upper panels} of Fig.~\ref{fig9}. The {\em middle and lower 
panels} show the variation of the Schechter parameters for each subsample as a function of group mass.
From these panels, it can be seen that there is clear influence of the large scale environment on
the resulting LF parameters, and on their variation with group mass. When analysing groups at the densest
large scale environments, we do not observe a significant variation as a function of group mass, 
showing almost a constant behaviour for both, the characteristic absolute
magnitude ($\sim -20.7$) and the faint end slope ($\sim -1.1$), and these
behaviours remains roughly unchanged moving from a small ($f=2$) to a large ($f=5$)
surrounding volume. On the other hand, a different behaviour is observed for the subsamples at
the lowest density surrounding environments. Regarding the characteristic absolute magnitude at the
smallest surrounding volumes ($f=2$), it can be seen that it is a decreasing function of mass,
varying from $-20.4$ to $-20.7$ ($\sim 0.3$ mag) over 1 order of group mass.
This particular trend becomes more and more pronounced when the surrounding volume is increased, finding 
the largest variation at the largest volume ($f=5$), from $-20.3$ to $-20.8$ ($\sim 0.5$ mag) over 1 order
of group mass.  A similar behaviour is observed for the faint end slope. At the smallest surrounding 
volume, we observe almost a constant value of $\sim-0.9$ as a function of group mass, which is a 
significantly shallower slope than that obtained in the densest subsample. When the surrounding volume
is increased, the $\alpha$-$\cal{M}$ relation, for the lowest density subsamples, becomes a 
decreasing function of mass, varying from $-0.85$ to $-1.1$ 
($\sim 0.25$) over 1 order in group mass, for the largest volume analysis. 

%%%%%%%%%%%%%%%%%%%%%%%%%%%%%%%%%%%%%%%%%%%%%%%%%%%%%%%%%%%%%%%%%%%%%%%%%%%%%%%
%%%%%%%%%%%%%%%%%%%%%%%%%%%%%%%%%%%%%%%%%%%%%%%%%%%%%%%%%%%%%%%%%%%%%%%%%%%%%%%
%%%%%%%%%%%%%%%%%%%%%%%%%%%%%%%%%%%%%%%%%%%%%%%%%%%%%%%%%%%%%%%%%%%%%%%%%%%%%%%
%%%%%%%%%%%%%%%%%%%%%%%%%%%%%%%%%%%%%%%%%%%%%%%%%%%%%%%%%%%%%%%%%%%%%%%%%%%%%%%
\section{Discussion}

In the last few years, the mass dependence of the LF of galaxies in groups has been
analysed by several authors (e.g \citealt{zmm06,robot10}). Our findings for the overall LF
are in agreement with the trends found in those previous works, but the use of a
larger sample of groups allowed us an important statistical improvement.
We have been able to disentangle the dependence of the LF on
galaxy types, the location of galaxies within groups
and the large scale environments of groups. 

Our results strongly suggest that the population of red spheroids is 
the most related to the group environment.
Among the different galaxy types, only red spheroids have LF that strongly correlates with group mass.
On the other hand, the late types (blue and red), have LFs that are independent of mass.
The mass dependence of the LF of red spheroids and their increasing fraction with mass are the
responsible for the mass dependence of the overall LF of galaxies in groups.
There are several physical mechanisms proposed in the literature to explain the presence of
red spheroids in groups, some of them are more effective in higher mass systems,
which may explain the increasing fraction of red spheroids with group mass.
For instance, in high mass groups, strangulation (e.g \citealt{larson80,balogh00}) 
and ram pressure (e.g. \citealt{gunn72,abadi}) can transform and quench star formation,
which, in turn, affect primarily galaxy colour, while galaxy harassment transforms disks
into spheroids (e.g. \citealt{farouki81,moore96}). Regardless the masses of the systems,
galaxy major mergers (e.g. \citealt{toomre72,hopkins08}) combined with a mechanism to prevent
subsequent star formation (such as AGN feedback, \citealt{bower06,croton06}) can be responsible
for the formation of red spheroids in groups of any masses.

Our work raises another interesting point. We find a clear indication of 
luminosity segregation, since galaxies in the inner parts
have brighter characteristic magnitudes than their counterparts in the outer regions. 
This agrees with the well-known fact that
the most luminous galaxies exist preferentially in the densest regions of
the Universe (e.g \citealt{davis88,hamilton88,loveday95,benoist99,zehavi05,deng07}).
Moreover, the luminosity segregation becomes more pronounced for massive systems as a consequence
of a strong brightening of $M^{\ast}$ with the group mass for galaxies in the inner parts.
These facts resemble the results of \citet{skibba07} in which, by testing the halo model
predictions in groups of galaxies, they found that the luminosity of the central objects increases with halo
mass, while non central galaxies show almost no mass dependence.    

Regarding the connection between the large scale group surrounding environment and the LF,
high density regions show brighter and steeper values
than the observed at low density regions, for both, $M_{\ast}$ and $\alpha$ respectively.
A similar result was observed by \citet{tempel10} when analysing galaxy luminosities in the
SDSS DR7 at different global density environments. They found that $M_{\ast}$ clearly becomes brighter
from voids to superclusters, however, they do not observe any variation in $\alpha$.
Furthermore, an interesting result arises in our work when analysing the variation of 
the LF parameters with group mass, since two distinct behaviours are observed: 
meanwhile galaxies in groups within high density regions have LF parameters that exhibit
almost no changes with group mass, LF of galaxies in groups 
inhabiting low density regions experience a significant variation with mass.
Our results suggest a plausible scenario for galaxy evolution in which both, large scale and local 
environments play important roles. 
There is evidence in the literature stating that groups in high density regions
formed earlier \citep{harker06}. This gives the galaxy members a longer time to evolve,
producing brighter bright galaxies ($M^{\ast}$), mostly through mergers, and also a larger number of 
faint galaxies ($\alpha$), affected by dynamical friction and with depleted gas reservoirs.
For these groups, the effect of large scale environment could be the main responsible for
galaxy evolution, with the group mass (local environment) playing a secondary role in the final result.
On the other hand, for groups inhabiting low density regions, the group mass 
plays a more important role in the observed galaxy luminosities.
For a fixed mass, it might be inferred that the difference in the formation time between 
groups in high and low density regions, is the key point to understand the differences in the LF,
i.e., we are observing different stages of the same galaxy evolution. 
However, the formation time can not be entirely blamed for all differences, since 
groups in high density regions may have accreted
a larger amount of material from their surroundings during their evolution, while
groups in low density regions have had  more limited access to fresh material. 
Thus, galaxy evolution in groups may follow different paths depending on where the groups
inhabit.

%%%%%%%%%%%%%%%%%%%%%%%%%%%%%%%%%%%%%%%%%%%%%%%%%%%%%%%%%%%%%%%%%%%%%%%%%%%%%%%
%%%%%%%%%%%%%%%%%%%%%%%%%%%%%%%%%%%%%%%%%%%%%%%%%%%%%%%%%%%%%%%%%%%%%%%%%%%%%%%
\section{Summary}
We have carried out an exhaustive analysis of the luminosity function of galaxies in groups.
This work was aimed to achieve a more complete and detailed analysis than previously done
by \citet{zmm06}. For this new study, we have identified groups of galaxies in the MGS of the
SDSS DR7, obtaining one of the largest group samples at present ($\sim 16,000$ groups with more than 4 members) 
that allowed us to obtain more refined and reliable statistics than in our previous work.

We have studied the group mass dependence of the galaxy LF under a number of different conditions: 
group identification algorithms, photometric bands,
galaxy types, galaxy locations within groups, and the large scale environment in which the
groups are positioned. We have found that the Schechter function is an appropriate description of the 
LF of galaxies in groups in all cases studied, thus our work relies on the analysis of the parameters
describing the characteristic absolute magnitude, $M^{\ast}$, and the faint end slope, $\alpha$,
and their dependence with group mass. In the Appendix table we quote the LF parameters
from all our analyses. The main results of our work can be summarised as follows:
\begin{itemize}
\item The LF is sensitive to the identification technique (single or double identification)
and to the value of the contour density contrast adopted. Our preferred choice, 
double identification plus high contour density contrast ($\delta\rho/\rho=200$),
gives LFs that have systematically brighter $M^{\ast}$ and steeper $\alpha$ compared to
single identifications and/or lower density contrast values. It should be emphasised that 
the latter options are prone to include a larger number of loose systems, whose existence and 
physical parameters are less reliable, thus biasing the LF results \citep{merchan02,mamon07}.  
\item The trends of the Schechter parameters as a function of group mass show very similar results 
regardless of the photometric band considered. Over the probed group mass range 
($12\le\log({\mathcal M}/M_{\odot} h^{-1})\le 15$), $M^{\ast}$ brightens in $\sim 0.6$ mag
and $\alpha$ decreases in $\sim 0.4$ for the $\gb$, $\rb$, $\ib$ and $\zb$ bands. 
For the $\ub$ band we find a brightening of $\sim 0.5$ mag in $M^{\ast}$ and a decrease of $\sim 0.6$
in $\alpha$. This different behaviour for the $\ub$ band is likely to be explained in terms of the 
spectral coverage of the $\ub$ band, that is closely related to the current
star formation, and the well known suppressed star formation rate experienced by galaxies in systems
\citep{zmm06}. 
\item When splitting galaxies into early and late types according to their concentration parameter,
or into blue and red according to their position in the colour-magnitude diagram, 
we obtain quite different results in the mass dependence of the LF, mainly for the faint end slope
of the different populations.
The $\alpha$ parameters of early and late types are decreasing functions of group mass, 
parallel one to another and shifted in $\Delta\alpha \sim 0.8$.  On the other hand, when classifying
galaxies according to colour, the $\alpha$ of the red population decreases with mass, but the
$\alpha$ of the blue population remains constant over the whole mass range.  
These differences reinforce the idea that there is no univocal relation between galaxy colour 
and morphology (e.g. \citealt{bundy10}).
Classifying galaxies using concentration index and colour, simultaneously, allows us to understand the nature of the
observed trends. We find that the blue-late and the red-late (passive disks) types, have LF parameters that do not
correlate with group mass, it is their relative fraction what changes with mass and determines
the trend of late types with mass. On the other hand, red-early type galaxies (red spheroids)
do have LF parameters that are strongly correlated with group mass. 
All mass dependencies of the LF parameters observed in this work can be understood by the contribution
of this single galaxy population.
\item Analysing galaxies in the inner and outer regions of groups, we find an indication of 
luminosity segregation in systems more massive than $\sim 10^{13} M_{\odot} h^{-1}$.
For these groups, galaxies in the inner parts have systematically brighter ($\sim 0.4$ mag)
characteristic magnitude. 
\item Finally, we analysed the possible influence of the large scale group surrounding environment
on the luminosities of group members.
It can be seen that, on average, the high density regions show brighter and steeper values
than the observed at low density regions, for both, $M_{\ast}$ and $\alpha$ respectively.
Galaxies in groups within high density regions have LF parameters that 
have almost no changes with group mass in contrast with galaxies in groups 
inhabiting low density regions.
\end{itemize}

The implications of our results should be addressed by theoretical and semianalytic models
in order to fully understand the physics behind the role played by the environment on galaxy evolution.

\section*{Acknowledgements}
We thank the anonymous referee for helpful suggestions.
We also thank Eugenia D\'iaz for useful comments.
This work has been partially supported with grants from Consejo 
Nacional de Investigaciones Cient\'\i ficas y T\'ecnicas de la Rep\'ublica 
Argentina (CONICET) and Secretar\'\i a de Ciencia y Tecnolog\'\i a de la 
Universidad de C\'ordoba.

Funding for the SDSS and SDSS-II has been provided by the Alfred P. Sloan Foundation, the Participating Institutions, the National Science Foundation, the U.S. Department of Energy, the National Aeronautics and Space Administration, the Japanese Monbukagakusho, the Max Planck Society, and the Higher Education Funding Council for England. The SDSS Web Site is http://www.sdss.org/.
The SDSS is managed by the Astrophysical Research Consortium for the Participating Institutions. The Participating Institutions are the American Museum of Natural History, Astrophysical Institute Potsdam, University of Basel, University of Cambridge, Case Western Reserve University, University of Chicago, Drexel University, Fermilab, the Institute for Advanced Study, the Japan Participation Group, Johns Hopkins University, the Joint Institute for Nuclear Astrophysics, the Kavli Institute for Particle Astrophysics and Cosmology, the Korean Scientist Group, the Chinese Academy of Sciences (LAMOST), Los Alamos National Laboratory, the Max-Planck-Institute for Astronomy (MPIA), the Max-Planck-Institute for Astrophysics (MPA), New Mexico State University, Ohio State University, University of Pittsburgh, University of Portsmouth, Princeton University, the United States Naval Observatory, and the University of Washington.

%%%%%%%%%%%%%%%%%%%%%%%%%%%%%%%%%%%%%%%%%%%%%%%%%%%%%%%%%%%%%%%%%%%%%%%%%%%%%%%
%%%%%%%%%%%%%%%%%%%%%%%%%%%%%%%%%%%%%%%%%%%%%%%%%%%%%%%%%%%%%%%%%%%%%%%%%%%%%%%

\label{lastpage}

\appendix
\section{Table}
In the following table we quote the STY best-fitting Schechter parameters for different subsamples of galaxies
in groups used in this work. 

\renewcommand{\tabcolsep}{0.05cm}
\renewcommand{\arraystretch}{0.7}

\begin{sidewaystable*}[h]
% \begin{center}
\centering
\begin{tabular}{c|cccc|c|cccc|c|cccc|c|cccc|c}
\cline{2-20}
\cline{2-20}
\vline & {\sz Mass range\footnotemark[1]} & {\sz $N_{\rm gal}$} & {\sz $M^{\ast}-5\log(h)$} & {\sz $\alpha$} & 
\vline & {\sz Mass range\footnotemark[1]} & {\sz $N_{\rm gal}$} & {\sz $M^{\ast}-5\log(h)$} & {\sz $\alpha$} & 
\vline & {\sz Mass range\footnotemark[1]} & {\sz $N_{\rm gal}$} & {\sz $M^{\ast}-5\log(h)$} & {\sz $\alpha$} & 
\vline & {\sz Mass range\footnotemark[1]} & {\sz $N_{\rm gal}$} & {\sz $M^{\ast}-5\log(h)$} & {\sz $\alpha$} & \vline \\\cline{2-20}
%%%%%%%%%%%%%%%%%%%%%%%%%%%%%%%%%%%%%%%%%% 
\vline & \multicolumn{4}{|c|}{\sz $14.50 \leq r \leq 17.76$} & 
\vline & \multicolumn{4}{|c|}{\sz $r$ band - Early} & 
\vline & \multicolumn{4}{|c|}{\sz $r$ band - Red \& Late} & 
\vline & \multicolumn{4}{|c|}{\sz $r$ band - $\rho_2$ low} & \vline \\\cline{2-20}
%-------------------------------------------------------------------------------------------------------------------------------------------------
\vline & {\sz $12.00-12.58$} & {\sz 7479} & {\sz $-20.61\pm0.05$} & {\sz $-1.04\pm0.03$} & 
\vline & {\sz $12.00-12.58$} & {\sz 3322} & {\sz $-20.48\pm0.06$} & {\sz $-0.49\pm0.05$} & 
\vline & {\sz $12.00-12.66$} & {\sz 1621} & {\sz $-20.41\pm0.15$} & {\sz $-1.61\pm0.08$} & 
\vline & {\sz $12.00-12.83$} & {\sz 7187} & {\sz $-20.43\pm0.05$} & {\sz $-0.87\pm0.04$} & \vline \\
%-------------------------------------------------------------------------------------------------------------------------------------------------
\vline & {\sz $12.58-12.84$} & {\sz 7998} & {\sz $-20.71\pm0.05$} & {\sz $-1.09\pm0.03$} & 
\vline & {\sz $12.58-12.84$} & {\sz 3815} & {\sz $-20.58\pm0.05$} & {\sz $-0.51\pm0.04$} & 
\vline & {\sz $12.66-12.94$} & {\sz 1785} & {\sz $-20.24\pm0.12$} & {\sz $-1.54\pm0.08$} & 
\vline & {\sz $12.83-13.16$} & {\sz 6047} & {\sz $-20.53\pm0.06$} & {\sz $-0.87\pm0.06$} & \vline \\
%-------------------------------------------------------------------------------------------------------------------------------------------------
\vline & {\sz $12.84-13.02$} & {\sz 8395} & {\sz $-20.65\pm0.04$} & {\sz $-1.01\pm0.03$} & 
\vline & {\sz $12.84-13.02$} & {\sz 4150} & {\sz $-20.53\pm0.05$} & {\sz $-0.44\pm0.05$} & 
\vline & {\sz $12.94-13.14$} & {\sz 1977} & {\sz $-20.35\pm0.12$} & {\sz $-1.65\pm0.09$} & 
\vline & {\sz $13.16-13.46$} & {\sz 3329} & {\sz $-20.58\pm0.09$} & {\sz $-0.89\pm0.08$} & \vline \\
%-------------------------------------------------------------------------------------------------------------------------------------------------
\vline & {\sz $13.02-13.18$} & {\sz 8681} & {\sz $-20.74\pm0.05$} & {\sz $-1.07\pm0.04$} & 
\vline & {\sz $13.02-13.18$} & {\sz 4619} & {\sz $-20.56\pm0.05$} & {\sz $-0.48\pm0.06$} & 
\vline & {\sz $13.14-13.33$} & {\sz 2048} & {\sz $-20.29\pm0.10$} & {\sz $-1.70\pm0.06$} & 
\vline & {\sz $>13.46$} & {\sz 1000} & {\sz $-20.68\pm0.16$} & {\sz $-0.98\pm0.15$} & \vline \\\cline{17-20}
%-------------------------------------------------------------------------------------------------------------------------------------------------
\vline & {\sz $13.18-13.33$} & {\sz 9277} & {\sz $-20.74\pm0.05$} & {\sz $-1.07\pm0.03$} & 
\vline & {\sz $13.18-13.33$} & {\sz 4970} & {\sz $-20.61\pm0.05$} & {\sz $-0.52\pm0.05$} & 
\vline & {\sz $13.33-13.52$} & {\sz 2191} & {\sz $-20.35\pm0.06$} & {\sz $-1.70\pm0.03$} & 
\vline & \multicolumn{4}{|c|}{\sz $r$ band - $\rho_2$ high} & \vline \\\cline{17-20}
%-------------------------------------------------------------------------------------------------------------------------------------------------
\vline & {\sz $13.33-13.49$} & {\sz 9538} & {\sz $-20.79\pm0.04$} & {\sz $-1.10\pm0.04$} & 
\vline & {\sz $13.33-13.49$} & {\sz 5312} & {\sz $-20.69\pm0.05$} & {\sz $-0.58\pm0.06$} & 
\vline & {\sz $13.52-13.73$} & {\sz 2235} & {\sz $-20.34\pm0.06$} & {\sz $-1.70\pm0.03$} & 
\vline & {\sz $12.00-12.83$} & {\sz 2360} & {\sz $-20.78\pm0.12$} & {\sz $-1.16\pm0.08$} & \vline \\
%-------------------------------------------------------------------------------------------------------------------------------------------------
\vline & {\sz $13.49-13.64$} & {\sz 10004} & {\sz $-20.88\pm0.05$} & {\sz $-1.16\pm0.04$} & 
\vline & {\sz $13.49-13.64$} & {\sz 5814} & {\sz $-20.73\pm0.05$} & {\sz $-0.58\pm0.06$} & 
\vline & {\sz $13.73-14.02$} & {\sz 2344} & {\sz $-20.29\pm0.06$} & {\sz $-1.70\pm0.03$} & 
\vline & {\sz $12.83-13.16$} & {\sz 3183} & {\sz $-20.61\pm0.10$} & {\sz $-1.05\pm0.07$} & \vline \\
%-------------------------------------------------------------------------------------------------------------------------------------------------
\vline & {\sz $13.64-13.84$} & {\sz 10473} & {\sz $-20.99\pm0.04$} & {\sz $-1.27\pm0.04$} & 
\vline & {\sz $13.64-13.84$} & {\sz 6368} & {\sz $-20.89\pm0.05$} & {\sz $-0.76\pm0.05$} & 
\vline & {\sz $>14.02$} & {\sz 2171} & {\sz $-20.40\pm0.05$} & {\sz $-1.70\pm0.01$} & 
\vline & {\sz $13.16-13.46$} & {\sz 5296} & {\sz $-20.80\pm0.07$} & {\sz $-1.14\pm0.06$} & \vline \\\cline{12-15}
%-------------------------------------------------------------------------------------------------------------------------------------------------
\vline & {\sz $13.84-14.10$} & {\sz 11348} & {\sz $-21.03\pm0.04$} & {\sz $-1.28\pm0.04$} & 
\vline & {\sz $13.84-14.10$} & {\sz 7207} & {\sz $-21.00\pm0.04$} & {\sz $-0.87\pm0.05$} & 
\vline & \multicolumn{4}{|c|}{\sz $r$ band - Blue \& Late} & 
\vline & {\sz $>13.46$} & {\sz 10007} & {\sz $-20.89\pm0.06$} & {\sz $-1.21\pm0.04$} & \vline \\\cline{12-20}
%-------------------------------------------------------------------------------------------------------------------------------------------------
\vline & {\sz $>14.10$}     & {\sz 12113} & {\sz $-21.18\pm0.04$} & {\sz $-1.38\pm0.04$} & 
\vline & {\sz $>14.10$}     & {\sz 8298} & {\sz $-21.18\pm0.04$} & {\sz $-1.03\pm0.05$} & 
\vline & {\sz $12.00-12.66$} & {\sz 3586} & {\sz $-20.25\pm0.07$} & {\sz $-1.20\pm0.05$} & 
\vline & \multicolumn{4}{|c|}{\sz $r$ band - $\rho_3$ low} & \vline \\\cline{2-10}\cline{17-20}
%%%%%%%%%%%%%%%%%%%%%%%%%%%%%%%%%%%%%%%%%% 
\vline & \multicolumn{4}{|c|}{\sz $16.50 \leq u \leq 19.00$} & 
\vline & \multicolumn{4}{|c|}{\sz $r$ band - Late} & 
\vline & {\sz $12.66-12.94$} & {\sz 3471} & {\sz $-20.21\pm0.07$} & {\sz $-1.22\pm0.06$} & 
\vline & {\sz $12.00-12.83$} & {\sz 5151} & {\sz $-20.48\pm0.07$} & {\sz $-0.91\pm0.06$} & \vline \\\cline{2-10}
%-------------------------------------------------------------------------------------------------------------------------------------------------
\vline & {\sz $12.00-12.58$} & {\sz 4379} & {\sz $-18.00\pm0.07$} & {\sz $-1.11\pm0.06$} & 
\vline & {\sz $12.00-12.58$} & {\sz 4157} & {\sz $-20.33\pm0.07$} & {\sz $-1.32\pm0.05$} & 
\vline & {\sz $12.94-13.14$} & {\sz 3196} & {\sz $-20.20\pm0.07$} & {\sz $-1.17\pm0.06$} & 
\vline & {\sz $12.83-13.16$} & {\sz 3406} & {\sz $-20.54\pm0.09$} & {\sz $-0.91\pm0.07$} & \vline \\
%-------------------------------------------------------------------------------------------------------------------------------------------------
\vline & {\sz $12.58-12.84$} & {\sz 4301} & {\sz $-17.98\pm0.07$} & {\sz $-1.14\pm0.07$} & 
\vline & {\sz $12.58-12.84$} & {\sz 4183} & {\sz $-20.19\pm0.07$} & {\sz $-1.34\pm0.05$} & 
\vline & {\sz $13.14-13.33$} & {\sz 3259} & {\sz $-20.29\pm0.07$} & {\sz $-1.28\pm0.06$} & 
\vline & {\sz $13.16-13.46$} & {\sz 3213} & {\sz $-20.59\pm0.09$} & {\sz $-0.92\pm0.07$} & \vline \\
%-------------------------------------------------------------------------------------------------------------------------------------------------
\vline & {\sz $12.84-13.02$} & {\sz 4206} & {\sz $-18.01\pm0.06$} & {\sz $-1.19\pm0.06$} & 
\vline & {\sz $12.84-13.02$} & {\sz 4245} & {\sz $-20.25\pm0.07$} & {\sz $-1.31\pm0.06$} & 
\vline & {\sz $13.33-13.52$} & {\sz 3133} & {\sz $-20.30\pm0.09$} & {\sz $-1.24\pm0.08$} & 
\vline & {\sz $>13.46$} & {\sz 3002} & {\sz $-20.76\pm0.11$} & {\sz $-1.07\pm0.09$} & \vline \\\cline{17-20}
%-------------------------------------------------------------------------------------------------------------------------------------------------
\vline & {\sz $13.02-13.18$} & {\sz 3932} & {\sz $-18.09\pm0.07$} & {\sz $-1.27\pm0.08$} & 
\vline & {\sz $13.02-13.18$} & {\sz 4062} & {\sz $-20.34\pm0.07$} & {\sz $-1.45\pm0.06$} & 
\vline & {\sz $13.52-13.73$} & {\sz 2964} & {\sz $-20.33\pm0.07$} & {\sz $-1.38\pm0.10$} & 
\vline & \multicolumn{4}{|c|}{\sz $r$ band - $\rho_3$ high} & \vline \\\cline{17-20}     
%-------------------------------------------------------------------------------------------------------------------------------------------------
\vline & {\sz $13.18-13.33$} & {\sz 3879} & {\sz $-18.16\pm0.07$} & {\sz $-1.36\pm0.08$} & 
\vline & {\sz $13.18-13.33$} & {\sz 4307} & {\sz $-20.33\pm0.07$} & {\sz $-1.46\pm0.06$} & 
\vline & {\sz $13.73-14.02$} & {\sz 2904} & {\sz $-20.36\pm0.07$} & {\sz $-1.27\pm0.11$} & 
\vline & {\sz $12.00-12.83$} & {\sz 2787} & {\sz $-20.71\pm0.11$} & {\sz $-1.09\pm0.08$} & \vline \\
%-------------------------------------------------------------------------------------------------------------------------------------------------
\vline & {\sz $13.33-13.49$} & {\sz 3431} & {\sz $-18.20\pm0.07$} & {\sz $-1.41\pm0.09$} & 
\vline & {\sz $13.33-13.49$} & {\sz 4227} & {\sz $-20.38\pm0.07$} & {\sz $-1.48\pm0.07$} & 
\vline & {\sz $>14.02$} & {\sz 2592} & {\sz $-20.36\pm0.06$} & {\sz $-1.22\pm0.13$} & 
\vline & {\sz $12.83-13.16$} & {\sz 4155} & {\sz $-20.68\pm0.09$} & {\sz $-1.04\pm0.06$} & \vline \\\cline{12-15}
%-------------------------------------------------------------------------------------------------------------------------------------------------
\vline & {\sz $13.49-13.64$} & {\sz 3043} & {\sz $-18.15\pm0.07$} & {\sz $-1.36\pm0.11$} & 
\vline & {\sz $13.49-13.64$} & {\sz 4190} & {\sz $-20.48\pm0.09$} & {\sz $-1.64\pm0.08$} & 
\vline & \multicolumn{4}{|c|}{\sz $r$ band - $r_p/r_{vir,p} < 0.4$} & 
\vline & {\sz $13.16-13.46$} & {\sz 5131} & {\sz $-20.79\pm0.09$} & {\sz $-1.15\pm0.06$} & \vline \\\cline{12-15}
%-------------------------------------------------------------------------------------------------------------------------------------------------
\vline & {\sz $13.64-13.84$} & {\sz 2590} & {\sz $-18.20\pm0.09$} & {\sz $-1.39\pm0.13$} & 
\vline & {\sz $13.64-13.84$} & {\sz 4106} & {\sz $-20.39\pm0.07$} & {\sz $-1.62\pm0.08$} & 
\vline & {\sz $12.00-12.58$} & {\sz 3259} & {\sz $-20.69\pm0.09$} & {\sz $-1.03\pm0.05$} & 
\vline & {\sz $>13.46$} & {\sz 7332} & {\sz $-20.86\pm0.06$} & {\sz $-1.19\pm0.05$} & \vline \\\cline{17-20}
%-------------------------------------------------------------------------------------------------------------------------------------------------
\vline & {\sz $13.84-14.10$} & {\sz 2193} & {\sz $-18.39\pm0.07$} & {\sz $-1.49\pm0.11$} & 
\vline & {\sz $13.84-14.10$} & {\sz 4144} & {\sz $-20.43\pm0.07$} & {\sz $-1.59\pm0.09$} & 
\vline & {\sz $12.58-12.84$} & {\sz 3660} & {\sz $-20.78\pm0.07$} & {\sz $-1.03\pm0.05$} & 
\vline & \multicolumn{4}{|c|}{\sz $r$ band - $\rho_4$ low} & \vline \\\cline{17-20}
%-------------------------------------------------------------------------------------------------------------------------------------------------
\vline & {\sz $>14.10$} & {\sz 1384} & {\sz $-18.46\pm0.07$} & {\sz $-1.70\pm0.06$} & 
\vline & {\sz $>14.10$} & {\sz 3831} & {\sz $-20.59\pm0.06$} & {\sz $-1.76\pm0.09$} & 
\vline & {\sz $12.84-13.02$} & {\sz 3894} & {\sz $-20.70\pm0.07$} & {\sz $-0.94\pm0.05$} & 
\vline & {\sz $12.00-12.83$} & {\sz 3895} & {\sz $-20.41\pm0.09$} & {\sz $-0.91\pm0.07$} & \vline \\\cline{2-11} 
%%%%%%%%%%%%%%%%%%%%%%%%%%%%%%%%%%%%%%%%%% 
\vline & \multicolumn{4}{|c|}{\sz $15.35 \leq g \leq 18.00$} & 
\vline & \multicolumn{4}{|c|}{\sz $r$ band - Red} & 
\vline & {\sz $13.02-13.18$} & {\sz 4118} & {\sz $-20.78\pm0.06$} & {\sz $-1.01\pm0.05$} & 
\vline & {\sz $12.83-13.16$} & {\sz 3460} & {\sz $-20.53\pm0.09$} & {\sz $-0.91\pm0.07$} & \vline \\\cline{2-10}
%-------------------------------------------------------------------------------------------------------------------------------------------------
\vline & {\sz $12.00-12.58$} & {\sz 5634} & {\sz $-19.60\pm0.06$} & {\sz $-0.97\pm0.05$} & 
\vline & {\sz $12.00-12.58$} & {\sz 3860} & {\sz $-20.61\pm0.06$} & {\sz $-0.82\pm0.04$} & 
\vline & {\sz $13.18-13.33$} & {\sz 4401} & {\sz $-20.85\pm0.06$} & {\sz $-1.06\pm0.05$} & 
\vline & {\sz $13.16-13.46$} & {\sz 3627} & {\sz $-20.64\pm0.10$} & {\sz $-0.97\pm0.08$} & \vline \\
%-------------------------------------------------------------------------------------------------------------------------------------------------
\vline & {\sz $12.58-12.84$} & {\sz 5830} & {\sz $-19.71\pm0.06$} & {\sz $-1.05\pm0.05$} & 
\vline & {\sz $12.58-12.84$} & {\sz 4547} & {\sz $-20.79\pm0.07$} & {\sz $-0.92\pm0.04$} & 
\vline & {\sz $13.33-13.49$} & {\sz 4504} & {\sz $-20.89\pm0.06$} & {\sz $-1.02\pm0.06$} & 
\vline & {\sz $>13.46$} & {\sz 4077} & {\sz $-20.79\pm0.09$} & {\sz $-1.08\pm0.08$} & \vline \\\cline{17-20}
%-------------------------------------------------------------------------------------------------------------------------------------------------
\vline & {\sz $12.84-13.02$} & {\sz 5983} & {\sz $-19.63\pm0.05$} & {\sz $-0.96\pm0.05$} & 
\vline & {\sz $12.84-13.02$} & {\sz 5007} & {\sz $-20.74\pm0.06$} & {\sz $-0.85\pm0.04$} & 
\vline & {\sz $13.49-13.64$} & {\sz 4751} & {\sz $-21.06\pm0.07$} & {\sz $-1.17\pm0.06$} & 
\vline & \multicolumn{4}{|c|}{\sz $r$ band - $\rho_4$ high} & \vline \\\cline{17-20}
%-------------------------------------------------------------------------------------------------------------------------------------------------
\vline & {\sz $13.02-13.18$} & {\sz 5851} & {\sz $-19.76\pm0.06$} & {\sz $-1.06\pm0.06$} & 
\vline & {\sz $13.02-13.18$} & {\sz 5522} & {\sz $-20.80\pm0.05$} & {\sz $-0.93\pm0.04$} & 
\vline & {\sz $13.64-13.84$} & {\sz 4920} & {\sz $-21.15\pm0.07$} & {\sz $-1.26\pm0.06$} & 
\vline & {\sz $12.00-12.83$} & {\sz 3607} & {\sz $-20.74\pm0.10$} & {\sz $-1.08\pm0.06$} & \vline \\
%-------------------------------------------------------------------------------------------------------------------------------------------------
\vline & {\sz $13.18-13.33$} & {\sz 5914} & {\sz $-19.75\pm0.05$} & {\sz $-1.04\pm0.05$} & 
\vline & {\sz $13.18-13.33$} & {\sz 5914} & {\sz $-20.85\pm0.05$} & {\sz $-0.98\pm0.04$} & 
\vline & {\sz $13.84-14.10$} & {\sz 4799} & {\sz $-21.21\pm0.09$} & {\sz $-1.31\pm0.07$} & 
\vline & {\sz $12.83-13.16$} & {\sz 4034} & {\sz $-20.64\pm0.09$} & {\sz $-1.00\pm0.06$} & \vline \\
%-------------------------------------------------------------------------------------------------------------------------------------------------
\vline & {\sz $13.33-13.49$} & {\sz 5733} & {\sz $-19.81\pm0.05$} & {\sz $-1.10\pm0.06$} & 
\vline & {\sz $13.33-13.49$} & {\sz 6353} & {\sz $-20.89\pm0.05$} & {\sz $-0.99\pm0.04$} & 
\vline & {\sz $>14.10$} & {\sz 3345} & {\sz $-21.24\pm0.09$} & {\sz $-1.29\pm0.08$} & 
\vline & {\sz $13.16-13.46$} & {\sz 4428} & {\sz $-20.74\pm0.09$} & {\sz $-1.11\pm0.06$} & \vline \\\cline{12-15}
%-------------------------------------------------------------------------------------------------------------------------------------------------
\vline & {\sz $13.49-13.64$} & {\sz 5564} & {\sz $-19.88\pm0.05$} & {\sz $-1.15\pm0.06$} & 
\vline & {\sz $13.49-13.64$} & {\sz 6819} & {\sz $-20.98\pm0.05$} & {\sz $-1.06\pm0.04$} & 
\vline & \multicolumn{4}{|c|}{\sz $r$ band - $r_p/r_{vir,p} > 0.4$} & 
\vline & {\sz $>13.46$} & {\sz 5944} & {\sz $-20.89\pm0.09$} & {\sz $-1.20\pm0.06$} & \vline \\\cline{12-20}
%-------------------------------------------------------------------------------------------------------------------------------------------------
\vline & {\sz $13.64-13.84$} & {\sz 5191} & {\sz $-20.00\pm0.06$} & {\sz $-1.26\pm0.07$} & 
\vline & {\sz $13.64-13.84$} & {\sz 7354} & {\sz $-21.13\pm0.05$} & {\sz $-1.21\pm0.05$} & 
\vline & {\sz $12.00-12.58$} & {\sz 4113} & {\sz $-20.50\pm0.07$} & {\sz $-1.03\pm0.04$} & 
\vline & \multicolumn{4}{|c|}{\sz $r$ band - $\rho_5$ low} & \vline \\\cline{17-20}
%-------------------------------------------------------------------------------------------------------------------------------------------------
\vline & {\sz $13.84-14.10$} & {\sz 5010} & {\sz $-20.05\pm0.05$} & {\sz $-1.30\pm0.06$} & 
\vline & {\sz $13.84-14.10$} & {\sz 8174} & {\sz $-21.19\pm0.05$} & {\sz $-1.25\pm0.04$} & 
\vline & {\sz $12.58-12.84$} & {\sz 4255} & {\sz $-20.64\pm0.07$} & {\sz $-1.12\pm0.04$} & 
\vline & {\sz $12.00-12.83$} & {\sz 3289} & {\sz $-20.33\pm0.09$} & {\sz $-0.85\pm0.07$} & \vline \\
%-------------------------------------------------------------------------------------------------------------------------------------------------
\vline & {\sz $>14.10$} & {\sz 3664} & {\sz $-20.16\pm0.05$} & {\sz $-1.32\pm0.07$} & 
\vline & {\sz $>14.10$} & {\sz 8886} & {\sz $-21.29\pm0.01$} & {\sz $-1.30\pm0.01$} & 
\vline & {\sz $12.84-13.02$} & {\sz 4410} & {\sz $-20.59\pm0.07$} & {\sz $-1.06\pm0.05$} & 
\vline & {\sz $12.83-13.16$} & {\sz 3488} & {\sz $-20.53\pm0.09$} & {\sz $-0.90\pm0.07$} & \vline \\\cline{2-10}   
%%%%%%%%%%%%%%%%%%%%%%%%%%%%%%%%%%%%%%%%%% 
\vline & \multicolumn{4}{|c|}{\sz $14.00 \leq i \leq 16.90$} & 
\vline & \multicolumn{4}{|c|}{\sz $r$ band - Blue} & 
\vline & {\sz $13.02-13.18$} & {\sz 4430} & {\sz $-20.66\pm0.07$} & {\sz $-1.12\pm0.06$} & 
\vline & {\sz $13.16-13.46$} & {\sz 3900} & {\sz $-20.73\pm0.09$} & {\sz $-1.03\pm0.07$} & \vline \\\cline{2-10}
%-------------------------------------------------------------------------------------------------------------------------------------------------
\vline & {\sz $12.00-12.58$} & {\sz 6442} & {\sz $-20.96\pm0.06$} & {\sz $-1.02\pm0.04$} & 
\vline & {\sz $12.00-12.58$} & {\sz 3619} & {\sz $-20.40\pm0.07$} & {\sz $-1.19\pm0.04$} & 
\vline & {\sz $13.18-13.33$} & {\sz 4664} & {\sz $-20.56\pm0.06$} & {\sz $-1.06\pm0.06$} & 
\vline & {\sz $>13.46$} & {\sz 4718} & {\sz $-20.80\pm0.07$} & {\sz $-1.11\pm0.06$} & \vline \\\cline{17-20}
%-------------------------------------------------------------------------------------------------------------------------------------------------
\vline & {\sz $12.58-12.84$} & {\sz 6854} & {\sz $-21.10\pm0.06$} & {\sz $-1.06\pm0.04$} & 
\vline & {\sz $12.58-12.84$} & {\sz 3451} & {\sz $-20.36\pm0.09$} & {\sz $-1.21\pm0.05$} & 
\vline & {\sz $13.33-13.49$} & {\sz 4665} & {\sz $-20.55\pm0.06$} & {\sz $-1.10\pm0.06$} & 
\vline & \multicolumn{4}{|c|}{\sz $r$ band - $\rho_5$ high} & \vline \\\cline{17-20}
%-------------------------------------------------------------------------------------------------------------------------------------------------
\vline & {\sz $12.84-13.02$} & {\sz 7343} & {\sz $-21.03\pm0.05$} & {\sz $-0.99\pm0.04$} & 
\vline & {\sz $12.84-13.02$} & {\sz 3388} & {\sz $-20.34\pm0.07$} & {\sz $-1.17\pm0.06$} & 
\vline & {\sz $13.49-13.64$} & {\sz 4662} & {\sz $-20.61\pm0.06$} & {\sz $-1.09\pm0.06$} & 
\vline & {\sz $12.00-12.83$} & {\sz 3971} & {\sz $-20.70\pm0.09$} & {\sz $-1.06\pm0.06$} & \vline \\
%-------------------------------------------------------------------------------------------------------------------------------------------------
\vline & {\sz $13.02-13.18$} & {\sz 7560} & {\sz $-21.11\pm0.05$} & {\sz $-1.04\pm0.04$} & 
\vline & {\sz $13.02-13.18$} & {\sz 3159} & {\sz $-20.39\pm0.07$} & {\sz $-1.21\pm0.06$} & 
\vline & {\sz $13.64-13.84$} & {\sz 4757} & {\sz $-20.78\pm0.06$} & {\sz $-1.25\pm0.07$} & 
\vline & {\sz $12.84-13.16$} & {\sz 3855} & {\sz $-20.65\pm0.09$} & {\sz $-1.01\pm0.07$} & \vline \\
%-------------------------------------------------------------------------------------------------------------------------------------------------
\vline & {\sz $13.18-13.33$} & {\sz 8126} & {\sz $-21.15\pm0.05$} & {\sz $-1.10\pm0.04$} & 
\vline & {\sz $13.18-13.33$} & {\sz 3363} & {\sz $-20.36\pm0.06$} & {\sz $-1.16\pm0.06$} & 
\vline & {\sz $13.84-14.10$} & {\sz 4844} & {\sz $-20.71\pm0.06$} & {\sz $-1.14\pm0.06$} & 
\vline & {\sz $13.16-13.46$} & {\sz 4214} & {\sz $-20.64\pm0.09$} & {\sz $-1.03\pm0.07$} & \vline \\
%-------------------------------------------------------------------------------------------------------------------------------------------------
\vline & {\sz $13.33-13.49$} & {\sz 8333} & {\sz $-21.21\pm0.05$} & {\sz $-1.12\pm0.05$} & 
\vline & {\sz $13.33-13.49$} & {\sz 3186} & {\sz $-20.39\pm0.07$} & {\sz $-1.23\pm0.08$} & 
\vline & {\sz $>14.10$} & {\sz 3637} & {\sz $-20.84\pm0.07$} & {\sz $-1.28\pm0.07$} & 
\vline & {\sz $>13.46$} & {\sz 5302} & {\sz $-20.86\pm0.09$} & {\sz $-1.20\pm0.06$} & \vline \\\cline{12-20}
%-------------------------------------------------------------------------------------------------------------------------------------------------
\vline & {\sz $13.49-13.64$} & {\sz 8784} & {\sz $-21.26\pm0.04$} & {\sz $-1.17\pm0.04$} & 
\vline & {\sz $13.49-13.64$} & {\sz 3185} & {\sz $-20.46\pm0.07$} & {\sz $-1.28\pm0.09$} & \vline \\
%-------------------------------------------------------------------------------------------------------------------------------------------------
\vline & {\sz $13.64-13.84$} & {\sz 9077} & {\sz $-21.41\pm0.05$} & {\sz $-1.30\pm0.05$} & 
\vline & {\sz $13.64-13.84$} & {\sz 3120} & {\sz $-20.44\pm0.07$} & {\sz $-1.28\pm0.10$} &  \vline \\
%-------------------------------------------------------------------------------------------------------------------------------------------------
\vline & {\sz $13.84-14.10$} & {\sz 9844} & {\sz $-21.49\pm0.05$} & {\sz $-1.36\pm0.05$} & 
\vline & {\sz $13.84-14.10$} & {\sz 3177} & {\sz $-20.45\pm0.07$} & {\sz $-1.22\pm0.10$} & \vline \\
%-------------------------------------------------------------------------------------------------------------------------------------------------
\vline & {\sz $>14.10$} & {\sz 10366} & {\sz $-21.64\pm0.04$} & {\sz $-1.46\pm0.05$} & 
\vline & {\sz $>14.10$} & {\sz 3243} & {\sz $-20.63\pm0.06$} & {\sz $-1.31\pm0.11$} & \vline \\\cline{2-10}  
%%%%%%%%%%%%%%%%%%%%%%%%%%%%%%%%%%%%%%%%%%
\vline & \multicolumn{4}{|c|}{\sz $14.20 \leq z \leq 17.25$} & 
\vline & \multicolumn{4}{|c|}{\sz $r$ band - Red \& Early} & \vline \\\cline{2-10}
%-------------------------------------------------------------------------------------------------------------------------------------------------
\vline & {\sz $12.00-12.58$} & {\sz 5901} & {\sz $-21.24\pm0.06$} & {\sz $-0.99\pm0.04$} & 
\vline & {\sz $12.00-12.66$} & {\sz 3327} & {\sz $-20.43\pm0.06$} & {\sz $-0.33\pm0.06$} & \vline \\
%-------------------------------------------------------------------------------------------------------------------------------------------------
\vline & {\sz $12.58-12.84$} & {\sz 6297} & {\sz $-21.38\pm0.06$} & {\sz $-1.04\pm0.04$} & 
\vline & {\sz $12.66-12.94$} & {\sz 4047} & {\sz $-20.48\pm0.05$} & {\sz $-0.31\pm0.05$} & \vline \\
%-------------------------------------------------------------------------------------------------------------------------------------------------
\vline & {\sz $12.84-13.02$} & {\sz 6762} & {\sz $-21.31\pm0.05$} & {\sz $-0.98\pm0.04$} & 
\vline & {\sz $12.94-13.14$} & {\sz 4759} & {\sz $-20.61\pm0.05$} & {\sz $-0.46\pm0.04$} & \vline \\
%-------------------------------------------------------------------------------------------------------------------------------------------------
\vline & {\sz $13.02-13.18$} & {\sz 6935} & {\sz $-21.39\pm0.06$} & {\sz $-1.02\pm0.05$} & 
\vline & {\sz $13.14-13.33$} & {\sz 5251} & {\sz $-20.61\pm0.05$} & {\sz $-0.46\pm0.05$} & \vline \\
%-------------------------------------------------------------------------------------------------------------------------------------------------
\vline & {\sz $13.18-13.33$} & {\sz 7403} & {\sz $-21.40\pm0.05$} & {\sz $-1.04\pm0.04$} & 
\vline & {\sz $13.33-13.52$} & {\sz 5891} & {\sz $-20.69\pm0.05$} & {\sz $-0.49\pm0.06$} & \vline \\
%-------------------------------------------------------------------------------------------------------------------------------------------------
\vline & {\sz $13.33-13.49$} & {\sz 7631} & {\sz $-21.45\pm0.05$} & {\sz $-1.07\pm0.05$} & 
\vline & {\sz $13.52-13.73$} & {\sz 6475} & {\sz $-20.81\pm0.04$} & {\sz $-0.63\pm0.05$} & \vline \\
%-------------------------------------------------------------------------------------------------------------------------------------------------
\vline & {\sz $13.49-13.64$} & {\sz 8016} & {\sz $-21.50\pm0.05$} & {\sz $-1.11\pm0.05$} & 
\vline & {\sz $13.73-14.02$} & {\sz 7556} & {\sz $-21.04\pm0.04$} & {\sz $-0.83\pm0.04$} & \vline \\
%-------------------------------------------------------------------------------------------------------------------------------------------------
\vline & {\sz $13.64-13.84$} & {\sz 8284} & {\sz $-21.65\pm0.05$} & {\sz $-1.25\pm0.06$} & 
\vline & {\sz $>14.02$} & {\sz 8765} & {\sz $-21.20\pm0.02$} & {\sz $-0.99\pm0.04$} & \vline \\\cline{7-10}
%-------------------------------------------------------------------------------------------------------------------------------------------------
\vline & {\sz $13.84-14.10$} & {\sz 9004} & {\sz $-21.70\pm0.05$} & {\sz $-1.30\pm0.06$} & \vline \\
%-------------------------------------------------------------------------------------------------------------------------------------------------
\vline & {\sz $>14.10$} & {\sz 9400} & {\sz $-21.85\pm0.05$} & {\sz $-1.41\pm0.06$} & \vline \\\cline{2-5}
\label{tab}
\end{tabular}
\footnotetext[1]{Units are $\log({\cal M}/(h^{-1}M_{\odot}))$}
% \end{center}
\end{sidewaystable*}

\label{lastpage}
\end{document}